# Численное исследование газодинамических процессов в условиях экспериментов по осаждению нанокластеров серебра

Роман. В. Мальцев

rmaltsev@gmail.com

**Аннотация.** В настоящей работе, усовершенствованный метод ПСМ применяется для численного анализа результатов экспериментов по осаждению нанокластеров серебра, активно проводимых в Институте теплофизики СО РАН последние годы в рамках исследований по разработке нанокомпозитных бактерицидных покрытий. В настоящей работе анализируются газодинамические эффекты в условиях экспериментов: параметры газовой струи, истекающей из источника паров серебра, движение паров серебра внутри источника и в истекающей струе при разных условиях, движение нанокластеров серебра различной массы (до 1024 атомов), формирование которых предполагается в различных местах источника. Расчеты позволили обнаружить сильную неизотермичность источника, которая была затем подтверждена экспериментально, определить место образования генерируемых источником зародышей нанокластеров (недогретый конфузор сопла) и, наконец, показать, что процесс образования зародышей нанокластеров внутри источника имеет гетерогенную природу. Рассчитанная в настоящей работе ширина струи нанокластеров массой 1024 атома, достигших подложки, даёт хорошее согласие с полученным в эксперименте профилем покрытия, нанесенного на подложку-мишень в форме узкой полосы нержавеющей стали, что дополнительно верифицирует численные исследования.

Данная работа основана на материалах кандидатской диссертации по специальности 01.02.05, в настоящий момент подготавливаемой к защите.

В данную работу включены только материалы одной из глав, посвященной применению метода прямого статистического моделирования для интерпретации экспериментальных данных.

# Оглавление

# 1. Введение

В данной работе продемонстрировано применение усовершенствованного метода ПСМ (1) для интерпретации результатов экспериментов.

В 2008-2010 году в лаборатории разреженных газов Института теплофизики СО РАН активно проводились экспериментальные исследования по осаждению бактерицидных пленок (2). На вращающийся барабан с подложкой-мишенью направлялись две струи: струя аргона с испарениями серебра создавалась реактором-испарителем, струя продуктов пиролиза гексафторпропиленоксида (прежде всего – тетрафторэтилен) создавалась реактором-подогревателем. Далее дано краткое изложение части полученых в этих экспериментах результатов.

Исследование полученных металлополимерных пленок на бактерицидность показало хорошие результаты даже при относительно небольшом процентном содержании серебра. Микроскопические исследования показали наличие на поверхности напыленных пленок наночастиц серебра диаметров 10-100 нм.

К сожалению, ограниченная экспериментальная база, хоть и позволяла в широких пределах варьировать условия осаждения и изучать их влияние на свойства напыления (прежде всего бактерицидные). Кроме того, конструкция реактора-испарителя серебра практически лишала возможности наблюдать происходящие внутри него процессы.

Эксперименты показали зависимость бактерицидности пленок от характерного размера наночастиц серебра, образовывающихся на поверхности, который, в свою очередь, зависел от условий в реакторе-испарителе (схематически изображённого на **Рис. 1-1**). Удивительно, что ожидаемого резкого увеличения бактерицидности наночастиц серебра при уменьшении их видимых размеров (и, соответственно, увеличении удельной поверхности) обнаружено не было. Всё это привело к стремлению описать и понять механизм образования наночастиц. Были выдвинуты несколько версий: на поверхности мишени из паров, в струе при гомогенной конденсации паров, на поверхности сверхзвуковой части сопла в результате реиспарения, в камере торможения реактора в результате гомогенной конденсации, на поверхностях камеры торможения реактора, на поверхности расплава.

Сделать заключение на основе анализа имевшихся экспериментальных данных было крайне затруднительно. Поэтому, экспериментальной группой была проведена серия дополнительных экспериментов по напылению серебра на покрытые углеродом медные сетки, которые затем исследовались методами просвечивающей электронной микроскопии. На сетках были обнаружены наночастицы двух видов: диаметром до 5 нм, и диаметром от 10 нм. Таким образом, кроме частиц в десятки нанометров, были дополнительно обнаружены гораздо меньшие частицы, разрешить которые растровой электронной микроскопией на отраженных электронах затруднительно. Кроме того, рост и коагуляция частиц на покрытой углеродом сетке затруднены, что говорит не в пользу версии роста частиц на поверхности мишени.

Использование метода ПСМ, адаптированного для нанотехнологических задач динамики разреженных газов, позволяет получить дополнительную информацию о возможных процессах при осаждении. Так, метод позволяет рассчитать как пространственное распределение плотности и потока паров атомарного серебра, так и оценить движение кластеров и даже небольших наночастиц в потоке несущего газа. Сравнение макропараметров для разных режимов осаждения (прежде всего, разных давлений аргона) может дать дополнительную информацию, полезную для интерпретации результатов экспериментов.

В настоящей работе внимание уделено только источнику серебряных паров.

Цель последующего численного исследования: прояснить механизм и место образования напыляемых кластеров в экспериментальной установке, определить параметры кластерной струи на подлёте к мишени, а также выяснить, в каких случаях результаты экспериментов объясняются газодинамическими эффектами, а в каких – иными происходящими в установке процессами. Прояснить эти вопросы исключительно опытным путём экспериментальной группе не удалось. Прояснение механизма образования нанокластеров, несомненно, подскажет направление для дальнейшей оптимизации источника нанокластеров серебра и повышения его эффективности.

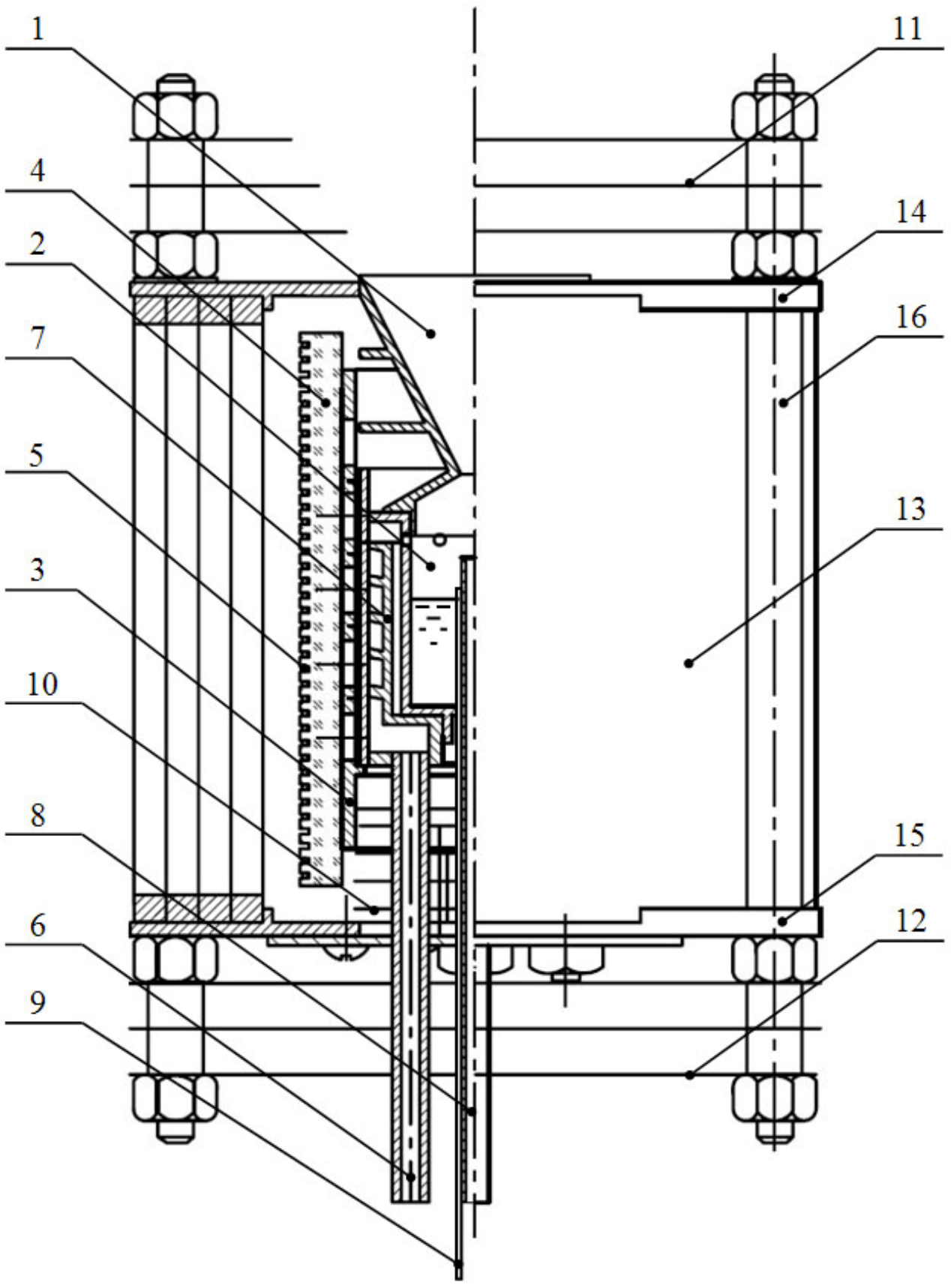

**Рис. 1-1**. Схема реактора-испарителя серебра. Аргон подаётся сначала во внешнюю камеру через трубку 6, проходит в ней через спиральный канал-теплообменник 7, и через расположенные по окружности отверстия попадает в верхнюю часть внутренней камеры (тигля) 2, где смешивается с парами серебра и затем истекает через сопло 1.

1 – сопло; 2 – тигель; 3 – каркас для керамики; 4 – керамический столбик; 5 – нихромовая проволока; 6 – трубка ввода аргона; 7 – витой канал-нагреватель аргона; 8 – трубка для измерения давления в тигле; 9 – термопарный кабель; 10, 11, 12, 13 – радиационные экраны; 14 – верхняя крышка источника; 15 – нижняя крышка источника; 16 – стойка.

# 2. Предварительные оценки условий в реакторе-испарителе

Характерная температура реактора в экспериментах составляет $T_0$ = 950°C (как правило, лежала в пределах 920÷980°C). Течение несущего газа в реакторе слабо меняется при изменении температуры в пределах 15 %, сильно меняется только давление насыщенных паров серебра. Давление в реакторе $P_0$ варьировалось в более широких пределах: от 1 до 3.5 мм рт. ст. Диаметр критического сечения сопла составляет 3 мм.

Теоретический расход аргона легко вычисляется по формуле: $Q = \rho_* c_* S_* = P_0 \sqrt{\frac{\gamma\mu}{RT_0}} \left(\frac{2}{\gamma+1}\right)^{\frac{1}{\gamma-1}+\frac{1}{2}} S_*$, где $\rho_*$– критическая плотность, $c_*$– критическая скорость, $S_*$ – площадь критического сечения сопла, $P_0$ и $T_0$ – давление и температура торможения соответственно, $R$ – универсальная газовая постоянная, $\mu$ – молярная масса, $\gamma$=5/3 – показатель адиабаты. При выбранных параметрах, теоретический расход аргона через реактор будет составлять 4.7 мг/с при максимальном давлении, или 0.16 н.л./мин. Реальный расход был в разы больше из-за негерметичности конструкции испарителя.

Предельная скорость потока – поступательная скорость, до которой разгонится газ, если вся его тепловая энергия перейдет в энергию направленного движения, т.е. при числе Маха $M \to \infty$ – вычисляется по формуле: $V_{max} = \sqrt{\frac{2\gamma}{\gamma-1}\frac{RT_0}{\mu}}$. В данных условиях $V_{max}$ = 1127 м/с.

Давление паров серебра в Па можно посчитать по эмпирической формуле: $P_{Ag} = 3.33682 \cdot 10^{17} \cdot T^{-1.8596} \cdot e^{-\frac{34881.7}{T}}$ (3). При различных температурах, давление насыщенных паров серебра составляет:

**Табл. 2.1**

| $T_0$, °C | 920 | 940 | 950 | 962 | 980 | 1100 |
|---|---|---|---|---|---|---|
| $P_{0\,Ag}$, мторр | 0.956 | 1.50 | 1.87 | 2.42 | 3.54 | 34.0 |
| $Q_{Ag}$, мг/час | 12.8 | 19.9 | 24.7 | 31.8 | 46.1 | 423 |

Температура 962°C близка к температуре плавления серебра 1235.1 K. Температура 1100°C позволила бы увеличить концентрацию паров серебра в 10 раз, но, к сожалению, недостижима в экспериментальной установке.

Если положить, что относительная концентрация серебра всюду постоянна, и примесь паров серебра не влияет на течение аргона, то, при данном $P_{Ag}$, массовый расход паров серебра через сопло при расходе аргона через сопло, равном теоретическому, составит $Q_{Ag} = P_{Ag}\sqrt{\frac{\gamma}{RT_0} \cdot \frac{\mu_{Ag}{}^2}{\mu_{Ar}}} \left(\frac{2}{\gamma+1}\right)^{\frac{1}{\gamma-1}+\frac{1}{2}} S_*$ . Значения $Q_{Ag}$ для разных температур так же приведены в **Табл. 2.1**. При максимальной температуре 980°C и минимальном давлении в реакторе 1 торр, массовая доля насыщенных паров серебра будет составлять менее 1%. Интересно сравнить расход паров серебра с максимально возможным потоком: $Q_{Ag-max} = P_{Ag}\sqrt{\frac{\mu_{Ag}}{2\pi RT_0}} S_{\text{исп}}$. Здесь $S_{\text{исп}}$ – площадь поверхности расплава (её диаметр 14 мм). В данном случае, отношение ${Q_{Ag}}/{Q_{Ag-max}}$ составляет около 14 %, т.е. чуть более 85 % серебра переконденсируется на поверхности расплава.

Длину свободного пробега серебро-серебро можно вычислить по формуле: $\lambda_{Ag} = \frac{kT_0}{\sqrt{2}\pi \cdot P_{Ag} \cdot d_{Ag}^2}$, что составит около 30 мм при $P_{Ag} = 3.54$ мторр и температуре 950°C. Это довольно большая величина при диаметре основной камеры реактора 13 мм. Для сравнения, длина свободного пробега аргон-аргон при минимальном давлении 1 торр составляет $\lambda_{Ar}$ = 0.27 мм. Впрочем, свободный пробег серебро-серебро в таких условиях не реализуется, из-за большой вероятности столкновений молекул паров серебра с аргоном.

Среднеквадратичное расстояние, которое проходит молекула серебряных паров за время $\delta t$ за счет диффузии, составляет: $L_{\delta t} = \sqrt{6D_{Ag-Ar}\delta t}$, где $D_{Ag-Ar}$ – коэффициент диффузии паров серебра в аргоне, его можно рассчитать по формулам из первой главы, используя параметры модели, приведенные ниже. $L_{\delta t} \sim \sqrt{\frac{1}{n_{Ar}}}$, где $n_{Ar}$ – концентрация аргона. Так, за время между столкновениями с аргоном, атомы паров серебра смещаются на $L_{Ag-Ar} \sim \frac{1}{n_{Ar}}$, что составляет 0.4 мм при давлении в реакторе 1 торр и 0.1 мм при давлении 3.5 торр. Что касается пробега между столкновениями серебро-серебро $L_{Ag-Ag} \sim \left(n_{Ar} n_{Ag}\right)^{-1/2}$, то эта величина меняется от 2.2 мм, при концентрации паров 3.5 мторр и давлении аргона 3.5 торр, и до 7.6 мм, при концентрации паров 1 мторр и давлении аргона 1 торр.

Среднее расстояние, которое проходит поток за время между столкновениями серебро-серебро, можно рассчитать по формуле $L = \frac{V}{n_{Ag} \cdot c_r \cdot \pi d_{Ag}^2}$. Здесь $V$ – скорость потока, $n_{Ag}$ – плотность паров серебра, $c_r$ – средняя относительная скорость серебро-серебро. Выразив $L$ через число Маха несущего газа, используя также изэнтропические формулы, получаем: $L \approx 1.33 \cdot \lambda_{Ag} \cdot M \left(1 + \frac{M^2}{3}\right)^{3/2}$.

Рассмотрим случай наибольшей концентрации серебра (3.5 мторр). Тогда $L \approx$ 40 мм при $M$ = 1, что сравнимо с размерами тигля вместе с соплом и, как будет показано далее, многократно превышает размеры сверхзвуковой зоны течения. Даже при наибольшем давлении насыщенных паров, формирование кластеров серебра в сверхзвуковой части струи маловероятно: из-за низкой концентрации паров серебра и высокой скорости, столкновения серебро-серебро в сверхзвуковой струе не успевают происходить. Также имеет смысл сравнить эту величину с диффузионным пробегом: $L \approx L_{Ag-Ag}$ при $M$=0.06. Из последнего можно заключить, что, в идеальных условиях, при $M$>0.06 превалирует конвективный перенос паров серебра, а при $M$<0.06 – начинает проявляться и диффузионный.

В области температур порядка $T_0$, зависимость давления насыщенных паров серебра от температуры имеет более крутой вид, чем изэнтропическая (адиабата) зависимость давления от температуры. Если считать относительную концентрацию серебра постоянной, как принято ранее, то пересыщение наступает при любых $M > 0$, и становится возможным образование кластеров в камере реактора. Однако, если по каким-то причинам концентрация паров серебра уменьшена, то пересыщение может наступать при довольно больших числах Маха. Обозначим за $\beta$ отношение истинной концентрации паров в камере торможения к максимальной, определяемой давлением насыщенных паров. Тогда, находя пересечение двух кривых давления (насыщенных паров и изэнтропы), можно вычислить точку пересыщения, а затем посчитать $L$. Для сравнения, можно вычислить характерный размер зоны пересыщения $R_S$ как диаметр, геометрическое число Маха для которого соответствует вычисленному числу Маха пересыщения. Данные приведены в следующей таблице:

**Табл. 2.2**. Оценки размеров зоны пересыщения при разных режимах.

| $\beta$ | Температура пересыщения, °C | Число Маха | $L$, мм | Характерный размер зоны пересыщения $R_S$, мм |
|---|---|---|---|---|
| 0.25 | 884 | 0.41 | 65 | 3.7 |
| 0.50 | 916 | 0.29 | 24 | 4.3 |
| 0.75 | 936 | 0.19 | 11 | 5.2 |
| 0.85 | 942 | 0.14 | 7.1 | 6.0 |
| 0.90 | 945 | 0.11 | 5.4 | 6.7 |
| 0.95 | 947 | 0.08 | 3.6 | 8.0 |
| 0.99 | 949 | 0.035 | 1.5 | 12.0 |

Вообще, сравнение $L$ с характерным диаметром для определенного числа Маха не является достаточным – при одном и том же характерном диаметре длина зоны пересыщения зависит от угла раствора конфузора $\theta$. Так, если применять конфузоры с малым углом раствора, то можно на порядок и более «растянуть» зону пересыщения и увеличить возможное число столкновений серебро-серебро, облегчив тем самым формирование кластеров. Поправочный коэффициент можно оценить по формуле: $\frac{R_S'}{R_S} = \frac{1}{4\sin\frac{\theta}{4}}$. Так, если вместо плавного конфузора использовать, наоборот, резкое изменение диаметра ($\theta$=180°), то характерный размер зоны пересыщения составит ≈ 0.35 от $R_S$. В данной конструкции реактора угол конфузора составлял примерно 113°, что дает оценку поправочного коэффициента ≈ 0.53.

Можно видеть, что уже при $\beta$ ≤ 0.95 формирование кластеров в потоке газа, сходящемся в конфузоре перед критическим сечением, становится крайне маловероятным, так как $L$ становится сравнимо с $R_S'$.

Причиной, по которой $\beta < 1$, может являться ограниченность диффузионного потока паров серебра сквозь слой покоящегося газа у испаряющейся поверхности. В результате, относительная концентрация паров серебра в потоке будет ниже, чем у зеркала расплава.

Для оценки $\beta$, а также более точного определения коэффициентов расхода, было проведено численное моделирование камеры реактора.

Для дальнейших оценок полезны и некоторые другие величины. Апроксимационная формула из (3) для температурной зависимости давления насыщенных паров серебра уже приведена ранее. Остальные величины, необходимые для расчета диаметра критического зародыша, также взяты из (3): радиус атома $r_a$ = 1.44 Å, объём в расчете на один атом $V_a$=17.14 Å$^3$, поверхностное натяжение $\gamma_\infty$ = 0.800 Н/м.

Полезными будут и теплофизические характеристики. Теплоёмкость серебра при комнатной температуре составляет 25.36 Дж/(моль·К), т.е. **3.05*R*** на каждый атом (*R* – универсальная газовая постоянная), что очень близко к классической теплоёмкости гармонического осциллятора, т.е. по 1*R* на каждое из 3 пространственных направлений. При температуре 800°С теплоёмкость повышается до **4.10*R*.**

Теплота плавления составляет 11.95 кДж/моль, т.е. соответствует теплоте нагрева серебра на **350К**. Теплота испарения составляет 254.1 кДж/моль, или **7 500 К**.

Интересно отметить, что поверхностной энергии наночастицы диаметром 1.3 нм (64 атома) соответствует теплота нагрева на 1000 К, и, при слиянии двух таких наночастиц в одну, они могут нагреться на 180 К, это составляет более половины теплоты плавления. Поверхностная энергия наночастицы диаметром 10 нм соответствует теплоте нагрева на 140 К, а слившись, две такие наночастицы нагреются всего на 30 К.

# 3. Параметры столкновительной модели и граничные условия

При изучении движения смеси аргона и паров серебра, из парных столкновений учитывались Ar-Ar и Ar-Ag. Столкновения Ag-Ag не учитывались, что оправдано низкой относительной концентрацией паров серебра (порядка 0.1 %). Модель столкновений молекул друг с другом – переменные мягкие сферы (VSS). Параметры модели VSS выбраны следующие:

**Табл. 3.1**. Параметры столкновительной модели

| Вид молекул | Молярная масса, а. е. | Диаметр, нм | ω | α |
|---|---|---|---|---|
| Ar | 40 | 0.328 | 0.65 | 1.28 |
| Ar-Ag | | 0.398 | 0.89 | 1.45 |
| Ag | 108 | 0.519 | 0.87 | 1.38 |

Параметры модели VSS рассчитаны на основе параметров модели Леннарда-Джонса. Для паров серебра последние (4) таковы: $d_{ref}$ = 2.644 Å, $T_{ref}$ = 4001 К. Для перекрестных столкновений использованы формулы осреднения (среднеарифметический диаметр, среднегеометрическая глубина ямы).

Для взаимодействия молекул со стенками реактора использовалась модель полного диффузного отражения.

Испаряющаяся поверхность моделировалась как источник частиц с полумаксвелловской функцией распределения, параметры которой соответствуют заданным $P_{Ag}$ и $T_0$. Возвращающиеся на поверхность частицы серебра полностью поглощались.

Несущий газ (аргон) конструктивно подается во внешнюю полость реактора, где прогревается до его температуры, после чего он втекает во внутреннюю полость реактора через 6 равномерно расположенных по окружности отверстий диаметром 3 мм. Решить трехмерную задачу, описывающую течение в реакторе целиком, не было возможности. Поэтому была использована осесимметричная модель, в которой газ подаётся в кольцевую щель. Оценить применимость модели можно следующим образом. При расчёте чисел Рейнольдса, коэффициентов сопротивления и т.п. часто используют эвристическую формулу учёта геометрии сечения, позволяющую определить некоторый характерный размер: $H_{**} = \frac{4S}{\Pi}$, где $S$ – плотность сечения, $\Pi$ – периметр. Для отверстия этот размер равен диаметру, для щели – удвоенной ширине. Суммарная площадь входных отверстий в 6 раз превышает площадь критического сечения. Соответственно, газ втекает в них с числом Маха M=0.094. Для того, чтобы получить ту же площадь и, соответственно, близкое число Маха, щель должна иметь ширину около 1 мм. При давлении 3.5 торр, число Рейнольдса для отверстия диаметром 3 мм составляет 5.54, а для щели шириной 1 мм, число Рейнольдса составит 3.56, что на 35% меньше. Поделать с этим ничего нельзя, так как изменение ширины щели приводит к почти кратному изменению скорости, не меняя число Рейнольдса. Тем не менее, числа Рейнольдса практически одного порядка, поэтому замена 6 отверстий щелью вполне допустима.

Щель является одновременно источником и поглотителем аргона. Параметры течения аргона подбираются такими, чтобы в камере торможения установилось нужное давление. Для паров серебра на поверхности щели задавалось зеркальное отражение, так как они не должны покидать границу расчетной области через щель.

Кроме того, в некоторых расчетах для более детального описания процессов в тигле моделировалась часть внешней камеры реактора.

# 4. Влияния буферного аргона на скорость испарения серебра

Влияние изучалось численно. Расчет течения во внутренней камере реактора производился при крайних значениях давления фонового газа (аргона) 1 мм рт. ст. и 3.5 мм рт. ст. Камера торможения в этих расчетах полагалась изотермической, т.е. конденсация серебра на стенках не учитывалась.

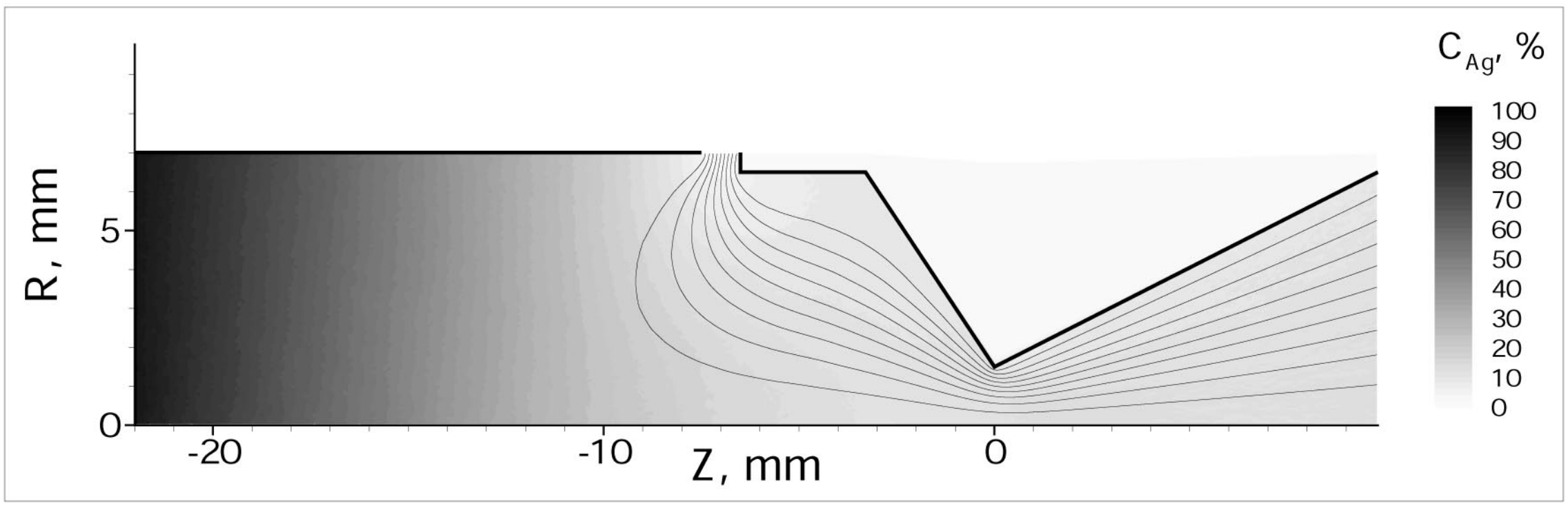

**Рис. 4-1**. Линии тока аргона и относительная мольная доля серебра в камере торможения реактора при большом расстоянии до поверхности серебра. За 100 % принята концентрация серебра у зеркала расплава.

На **Рис. 4-1** показаны линии тока несущего газа (аргона) и поле относительной концентрации паров серебра $\alpha$ для режима с давлением фонового газа 3.5 мм рт. ст. За 100 % принято значение относительной мольной доли паров серебра, соответствующей её теоретическому значению $P_{Ag}/P_0$ вблизи испаряющейся поверхности. Видно, что, по мере приближения к критическому сечению, относительная концентрация паров серебра падает, что связано с ограниченностью диффузионного потока паров серебра с поверхности вглубь потока.

Для сравнения, на **Рис. 4-2** показаны линии тока несущего газа и $\alpha$ при близком расстоянии до зеркала расплава. Можно видеть, что на этот раз линии тока проходят очень близко к расплаву серебра, эффективно унося пары, и концентрация паров серебра остается высокой во всем потоке.

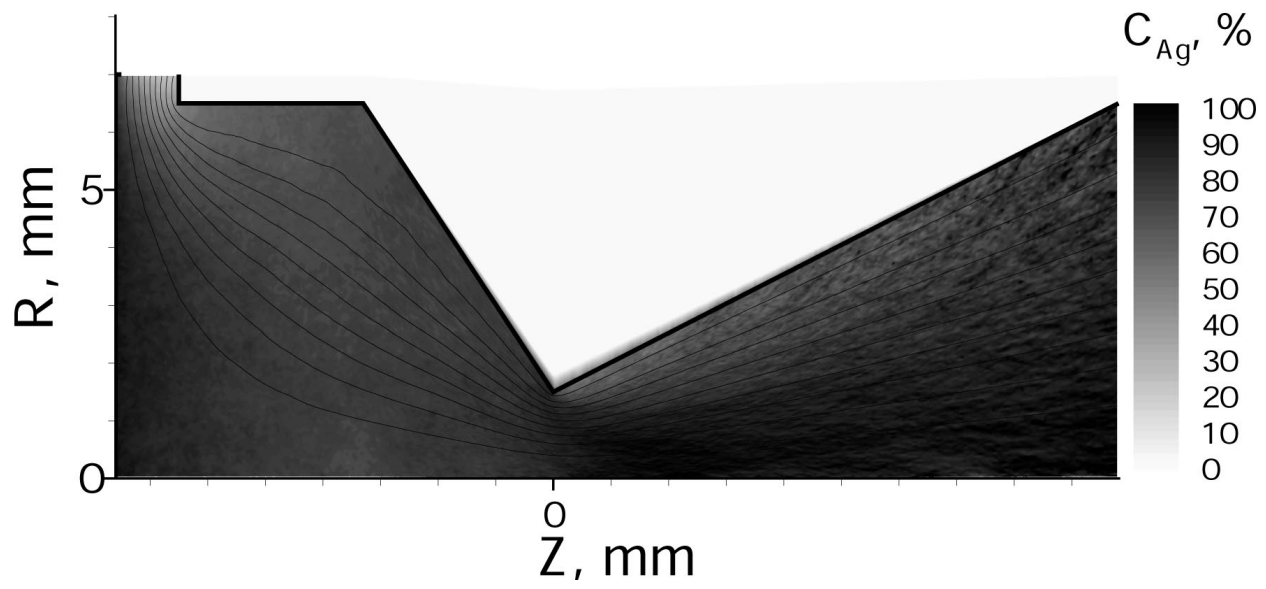

**Рис. 4-2**. Линии тока аргона и относительная мольная доля серебра в камере торможения реактора при малом расстоянии до поверхности серебра. За 100 % принята концентрация серебра у зеркала расплава.

В сводной **Табл. 4.1** приведены рассчитанные значения различных характеристик реактора при крайних режимах. Коэффициент расхода аргона – отношение вычисленного расхода к предельному теоретическому значению $Q_{Ar}$ – характеризует влияние вязкого сопротивления на пропускную способность сопла. Очевидно, увеличение давления аргона приводит к росту коэффициента расхода. Аргон «выдувает» пары серебра из реактора, направляя их на мишень, что и предопределило его использование в эксперименте. Коэффициент расхода паров серебра характеризует эффективность «выдувания» паров серебра аргоном и, в идеале ($\beta$ = 1 и отсутствие скольжения относительно аргона), равен коэффициенту расхода аргона. Так же приведена относительная концентрация паров серебра в критическом сечении, т.е. $\beta$. Её превышение над значениями, которые можно было бы ожидать из соотношения коэффициентов расхода, косвенно свидетельствует о явлении обогащения в сопле – вследствие диффузионных процессов пары серебра имеют меньшую скорость истечения, чем аргон, и, вследствие сохранения расхода, его относительная концентрация немного повышается. Эффективность разгона показывает отношение средней кинетической энергии серебра на срезе сопла к предельному значению $V_{max}$, когда вся тепловая энергия аргона переходит в энергию его направленного движения. Приведенные значения эффективности разгона служат лишь для ориентировки и не являются полностью достоверными, так как реально в вакуумной камере имеется фоновое давление, в то время как в расчете за соплом полагался глубокий вакуум. Из расчетов видно, что повышение давления фонового газа увеличивает эффективность разгона паров серебра и уменьшает сопротивление сопла. С другой стороны, повышенное давление значительно снижает скорость диффузии паров серебра сквозь столб аргона. Влияние этого фактора было бы менее существенным, если бы в реакторе реализовалось вихревое течение, эффективно перемешивающее газ. Но расчеты показали, что вихревых движений в диапазоне рабочих параметров не возникает (что не удивительно, так как числа Рейнольдса очень малы), а втекающий поток аргона проникает в сторону поверхности расплава всего на 2-3 мм. Варьируя расстояние между зеркалом серебряного расплава и отверстиями для подачи аргона, можно в широких пределах варьировать расход паров серебра. Однако, с экспериментальной точки зрения интереснее именно увеличивать расход серебра, а для этого необходимо максимально приблизить зеркало расплава к отверстиям подачи аргона.

**Табл. 4.1**. Производительность изотермического реактора по данным расчетов.

| | Расстояние от критического сечения до поверхности серебра 22 мм. | | Расстояние от критического сечения до поверхности серебра 7.6 мм. | |
|---|---|---|---|---|
| | Давление аргона 1 мм рт. ст. | Давление аргона 3.5 мм рт. ст. | Давление аргона 1 мм рт. ст. | Давление аргона 3.5 мм рт. ст. |
| Коэффициент расхода аргона | 70 % | 81 % | 68 % | 79 % |
| Коэффициент расхода паров серебра | 23 % | 8 % | 56 % | 62 % |
| Относительная концентрация паров серебра $\beta$ в критическом сечении по отношению к концентрации у поверхности расплава | 40 % | 11 % | 100 % | 85 % |
| Эффективность разгона паров серебра | 46 % | 63 % | 45 % | 62 % |

Выводы, полученные на основании проведенных расчетов, были использованы на практике для усовершенствования схемы эксперимента. Например, стало ясно, что уровень поверхности расплава должен располагаться как можно ближе к отверстиям, чтобы концентрация паров была максимальна.

Дополнительно, из данных можно сделать вывод, что практически во всем диапазоне расчетных параметров, конденсация (образование кластеров) в изотермической камере торможения реактора невозможна.

# 5. Оценка неизотермичности реактора-испарителя

При разработке реактора предполагалось его изотермичность. Однако, конструктивные ограничения – прежде всего, наличие сопла – способны привести к неравномерному распределению температур. Это очень важно, так как давление насыщенных паров сильно зависит от температуры.

Передача тепла излучением между двумя поверхностями: $q = \mathrm{C}\sigma S\left({T_1}^4 - {T_2}^4\right)$; где $\sigma$ – константа Стефана-Больцмана, С – коэффициент черноты.

Передача тепла через слой материала согласно закону Фурье: $q = \frac{\kappa S}{d}(T_1 - T_2)$; здесь $q$ – тепловой поток, $\kappa$ – коэффициент теплопроводности материала, $S$ – площадь слоя, $d$ – толщина слоя, $T_1$ и $T_2$ – температуры на границах слоя. Коэффициент теплопроводности для нержавеющей стали 12Х18Н10Т при температуре 950°С равен 27 Вт/(м·°К) (5).

При толщине стенок реактора 1 мм, неизотермичность по их толщине пренебрежимо мала, о чём говорит порядок величины $d_{\text{экв}} = \frac{\kappa}{4\sigma T^3} \approx 65\ \text{мм} \gg 1\ \text{мм}$. $d_{\text{экв}}$ – толщина слоя нержавеющей стали, при которой теплопередача через неё будет такой же, как между двумя излучающими пластинами с такой же (небольшой) разностью температур.

Рассмотрев вспомогательную модельную задачу о распределении температур вдоль тонкой пластины, нагреваемой с одного конца и охлаждаемой излучением, можно вычислить характерный размер градиента: $L_e = \sqrt{\frac{\kappa d_{\text{толщ}}}{8\mathrm{C}\sigma T^3}}$. Здесь $d_{\text{толщ}}$ – толщина пластины, $T$ – температура окружающего пространства. При коэффициенте черноты $\mathrm{C} = 1$, толщине пластины 1 мм и температуре 950°С, $L_e \approx 5.7\ \text{мм}$. Эта величина сравнима с размерами элементов конструкции реактора. С уменьшением коэффициента черноты, $L_e$ возрастает. Вообще, степень черноты нержавеющей стали при температуре 950°С составляет порядка $\mathrm{C} = 0.3$ (6), при которой $L_e \approx 10\ \text{мм}$. Однако, реальная степень черноты материала в реальных условиях работы конструкции может существенно отличаться от справочного значения.

Вывод: если тепловое излучение с разных сторон падает на элементы конструкции реактора неравномерно, то возможна существенная неравномерность температуры реактора.

В наихудших условиях в отношении изотермичности находится сопло реактора: так как через него в окружающее пространство истекает струя газа, нет возможности полностью окружить его тепловым экраном. С конструктивной точки зрения, сопло, включающее в себя конфузор и сверхзвуковую часть, выполнено как одна деталь из нержавеющей стали. Конфузор с помощью фаски стыкуется с цилиндрическим тиглем и фиксируется в нескольких точках точечной сваркой. Такой способ крепления ставит под вопрос надёжность теплового контакта между тиглем и соплом. Тем не менее, даже идеальный тепловой контакт не гарантирует изотермичности сопла: размеры одного только конфузора сравнимы с величиной $L_e$, а размеры сверхзвуковой части существенно её превышают. Кроме того, с точки зрения прогрева теплопроводностью, сопло имеет «узкое место» в критическом сечении, где поперечное сечение стенки сопла в 5 раз меньше по сравнению с поперечным сечением цилиндрического тигля.

Полагая для простоты сопло всё же изотермичным, будем характеризовать его «среднюю» температуру $T_{nzl}$ величиной $\alpha$, используя следующую формулу: $T_{nzl} = \left[(1-\alpha){T_2}^4 + \alpha {T_1}^4\right]^{1/4}$, где $T_1 = 300\ \text{K}$, $T_2 = 1223\ \text{K} = 950°\text{C}$ . С физической точки зрения, можно интерпретировать $\alpha$ как долю эффективной поверхности, обращённую в холодную область (оставшаяся часть обращена в область с температурой $T_2$).

График этой зависимости представлен на **Рис. 5-1**. Из рисунка видно, что уже при $\alpha = 0.25$, сопло может быть холоднее тигля на $85°\text{C}$. При $\alpha = 0.5$, отличие достигает почти $200°\text{C}$.

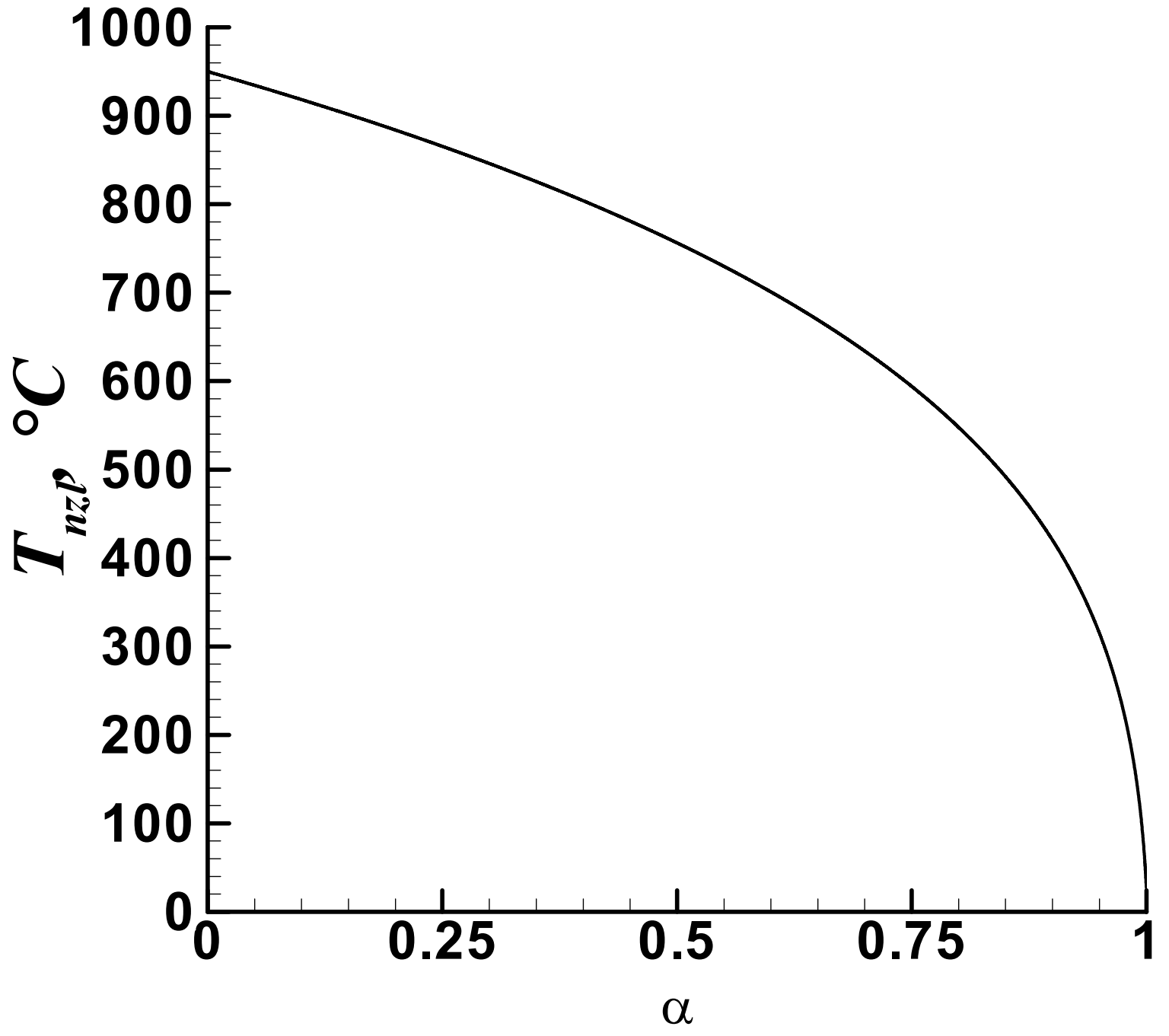

**Рис. 5-1**. Зависимость оценки характерной температуры сопла от доли поверхности, обращенной в окружающее холодное пространство.

Касательно изотермичности тигля, можно отметить следующее. Объём, занятый серебром, безусловно, можно считать изотермичным (при условии, что высота уровня составляет по крайней мере несколько миллиметров), так как серебро обладает высокой теплопроводностью. Температура стенки тигля между поверхностью расплава и отверстиями подачи аргона, по-видимому, тоже меняется слабо. Отверстия подачи аргона снижают эффективную площадь поперечного сечения стенки на 40%, что должно ослабить охлаждение тигля соплом. Если тигель обогревается снаружи равномерно, то не следует ожидать существенных температурных градиентов на поверхности тигля до места расположения отверстий.

Тепловой расчёт будет неполным без оценки теплового влияния потока несущего газа (аргона). При максимальном давлении, использовавшемся в реакторе (3.5 мм рт. ст.), при характерной температуре 950°C, теоретический предельный расход аргона через сопло с диаметром критического сечения 3 мм, составляет менее 5 мг/с. Таким образом, теплоёмкость (при постоянном давлении) количества аргона, истекающего через сопло в единицу времени, составляет порядка 0.0025 Вт/°К. Полное теплосодержание такого расхода газа (верхняя оценка энтальпии) составляет 3 Вт. Для сравнения, мощность нагревателя реактора в рабочем режиме составляет порядка 400 Вт. Несомненно, аргон не в состоянии оказать заметное влияние на тепловой баланс реактора.

Сделанные оценки использованы как весомое обоснование необходимости проведения эксперимента по измерению реальных температур сопла. Такой эксперимент был проведён в лаборатории экспериментальной группой, результаты приведены в таблице ниже:

**Табл. 5.1**. Результаты эксперимента по измерению температур сопла

| Давление в реакторе | 1.58 torr | 3.43 torr | 1.76 torr | 3.45 torr |
|---|---|---|---|---|
| Давление в вакуумной камере | 0.12 torr | 0.13 torr | 0.10 torr | 0.12 torr |
| Мощность нагревателя | 421 Вт | 424 Вт | 377 Вт | 378 Вт |
| Расход аргона | 0.24 н.л./мин | 0.55 н.л./мин | 0.24 н.л./мин | 0.55 н.л./мин |
| Температура серебра | 980 °C | 980 °C | 940 °C | 940 °C |
| Температура вблизи начала конфузора (вблизи стыка с тиглем) | Датчик вышел из строя | | | |
| Температура вблизи критического сечения сопла | 783 °C | 785 °C | Датчик вышел из строя | |
| Температура вблизи середины сверхзвуковой части сопла | 717 °C | 719 °C | 702 °C | 701 °C |
| Температура вблизи среза сверхзвукового сопла | Датчик вышел из строя | | | |

Измеренное превышение теоретического расхода аргона в 3.5 раза связано с негерметичностью реактора и канала подачи несущего газа. Расход аргона измерялся на входе в канал, давление газа в тигле измерялось через отдельный патрубок. К сожалению, в связи с выходом датчиков из строя в процессе эксперимента, данные оказались неполными.

Согласно экспериментальным данным, температура в критическом сечении на 197 °C отличается от температуры над поверхностью расплава, что хорошо соответствует значению $\alpha \approx 0.5$. Для середины сверхзвуковой части сопла хорошим приближением является $\alpha \approx 0.6$. Если приближенно положить, что характерной температуре конфузора соответствует $\alpha \approx 0.4$, то имеем оценку его температуры 880 °C.

Весьма полезным будет сравнить давление насыщенных паров серебра при разных температурах. Так, при температуре 980 °C давление насыщенных паров серебра составляет 3.54 миллиторр. При температуре 880 °C, давление насыщенных паров составляет около 10 % от исходного. При температуре 784 °C – уже 0.7 % от исходного, а при 718 °C – всего 0.1 %.

Следует уделить внимание и тепловым экранам установки. Тигель со всех сторон окружён 4 тепловыми экранами. Оценка сверху для температуры внешнего экрана – $\alpha = 0.8$, т.е. 550 °C. При такой температуре, давление паров снижено почти на 6 порядков по сравнению со значением у поверхности расплава, поэтому в расчетах оправданно пренебрегать отсутствием реиспарения паров с поверхности. Реально можно ожидать существенную неизотермичность тепловых экранов, однако, рассчитать профиль температуры внешнего теплового экрана весьма затруднительно.

Верхняя оценка для температуры подложки-мишени – 420 °C. Если подложка имеет сравнимый с тиглем (или бóльший) размер, и/или расположена достаточно далеко, её температура будет существенно ниже. Так, температура вращающегося барабана, на котором располагаются подложки в экспериментах по нанесению бактерицидных покрытий, оценена в 150 °C.

Сделанные оценки, свидетельствующие о сильной неизотермичности сопла, подтверждены экспериментально.

Изотермичность косвенно подтверждается также осмотром демонтированного тигля. Первые эксперименты проводились с использованием сплава серебра с 28 % содержанием меди. Последующие эксперименты проводились с чистым серебром, для чего был изготовлен ещё один тигель. Это дало возможность демонтировать и вскрыть для осмотра тигель, использовавшийся прежде. Фотографии поверхностей и тигля показы на **Рис. 5-2**.

На поверхности конфузора четко видна желтоватая шероховатая поверхность осажденного слоя вещества. Состав и происхождение тонкой поверхностной желтоватой пленки неизвестны. Тыльная сторона пленки имеет очень тёмный зелёный цвет. Под этим слоем обнаружен сравнительно толстый слой осажденного металла, покрытый сетью тонких складок. Наличие осажденного металла однозначно говорит о наличии существенного пересыщения пара в конфузоре и, как следствие, о сниженной температуре поверхности сопла. Неожиданным оказалось возвышение застывшей поверхности металла бугром при удалении от стенки тигля. Это свидетельствует о том, что жидкая поверхность расплавленного металла не смачивала стенки тигля. Стенки тигля выполнены из нержавеющей стали 12Х18Н10Т, которая смачивается как медью, так и серебром. По-видимому, смачиваемость исчезает в результате физического и/или химического взаимодействия поверхности тигля с посторонними веществами, такими как продукты термического распада вакуумного масла и примеси в баллоне с аргоном (прежде всего кислород).

А

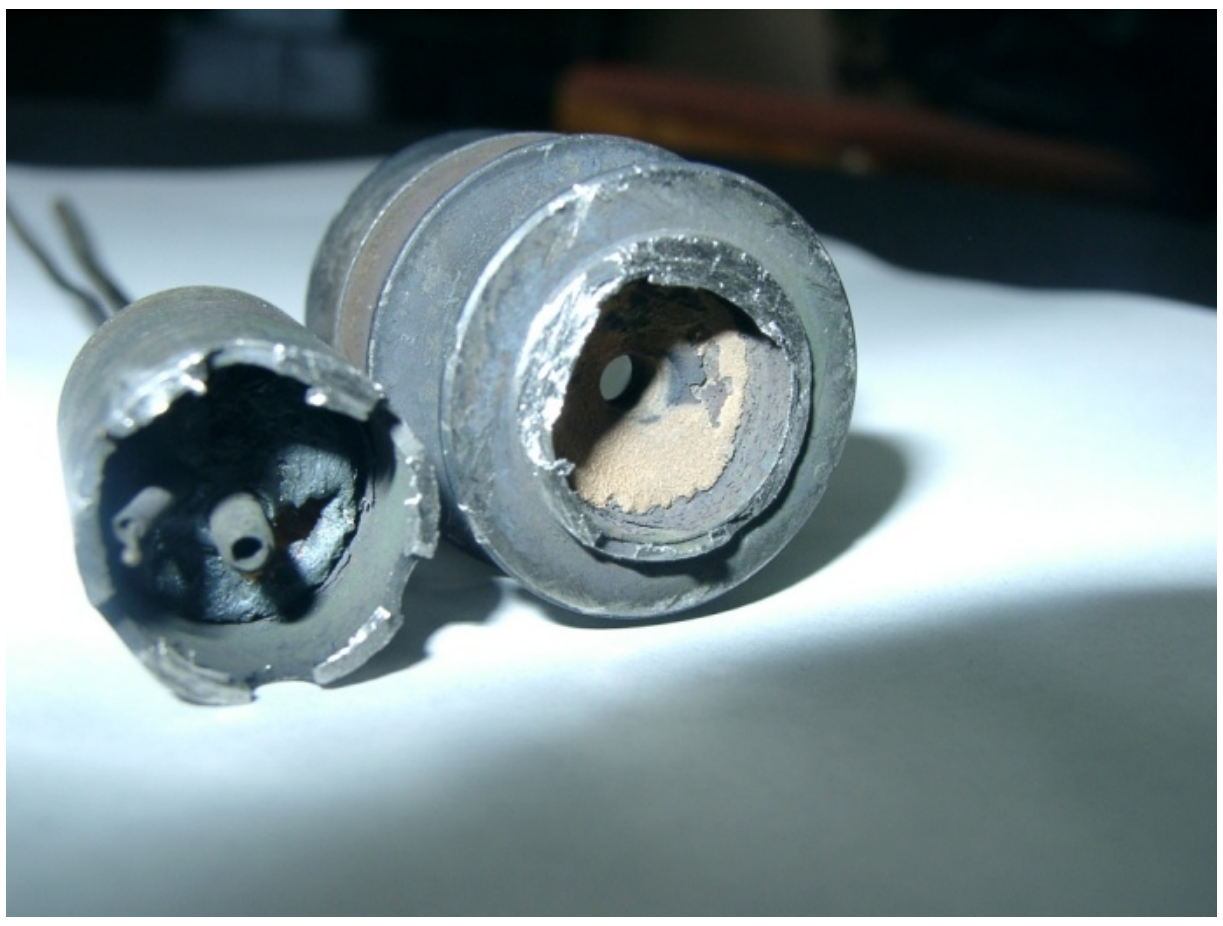

Б

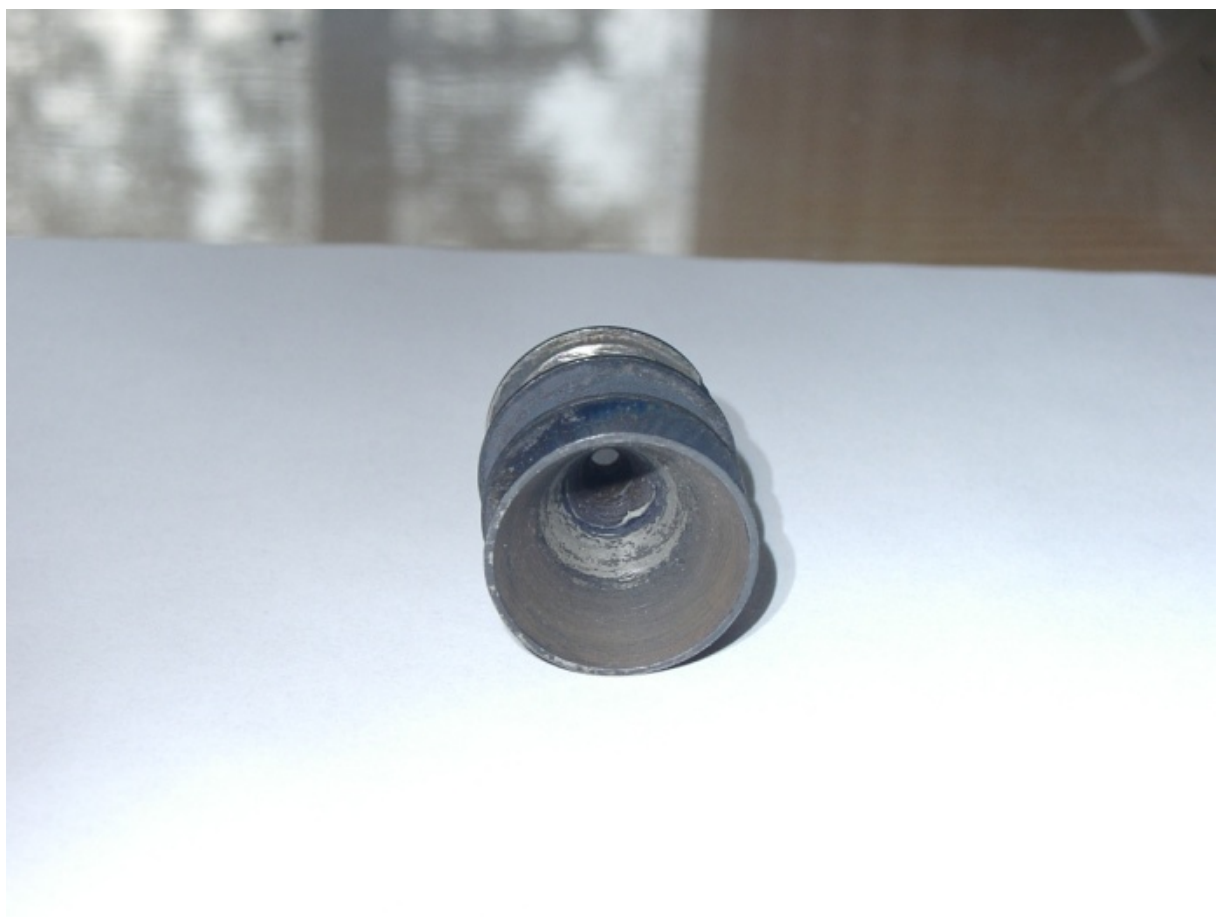

**Рис. 5-2**. Фотография вскрытого тигля (А) и диффузора сверхзвукового сопла (Б)

На фотографии диффузора сопла также виден серебристый след осажденного металла, располагающийся, однако, преимущественно в середине. Это также свидетельствует о конденсации пара на поверхности диффузора сверхзвукового сопла.

Факт конденсации серебра на диффузоре сверхзвукового сопла установить не трудно, так как серебряный след виден снаружи установки невооруженным глазом. Однако, наличие осаждённого вещества на конфузоре внутри источника для экспериментальной группы оказалось неожиданностью. При постановке экспериментов не возникало необходимости вскрывать источник, изначально это даже не было предусмотрено его конструкцией, поэтому внутренние поверхности визуальному контролю не подвергались. Только после того, как экспериментальная группа ознакомилась с изложенными выше результатами, показавшими несогласуемость расчетов изотермического источника с имевшимися данными экспериментов и поставила эксперимент, подтвердивший неизотремичность, возник повод вскрыть один из источников и изучить его изнутри.

Обнаружение осажденного вещества на стенках источника может свидетельствовать о вовлеченности гетерогенных процессов в образование наноклестеров, обнаруживаемых в экспериментах.

# 6. Оценка возможности гомогенной конденсации серебра в тигле

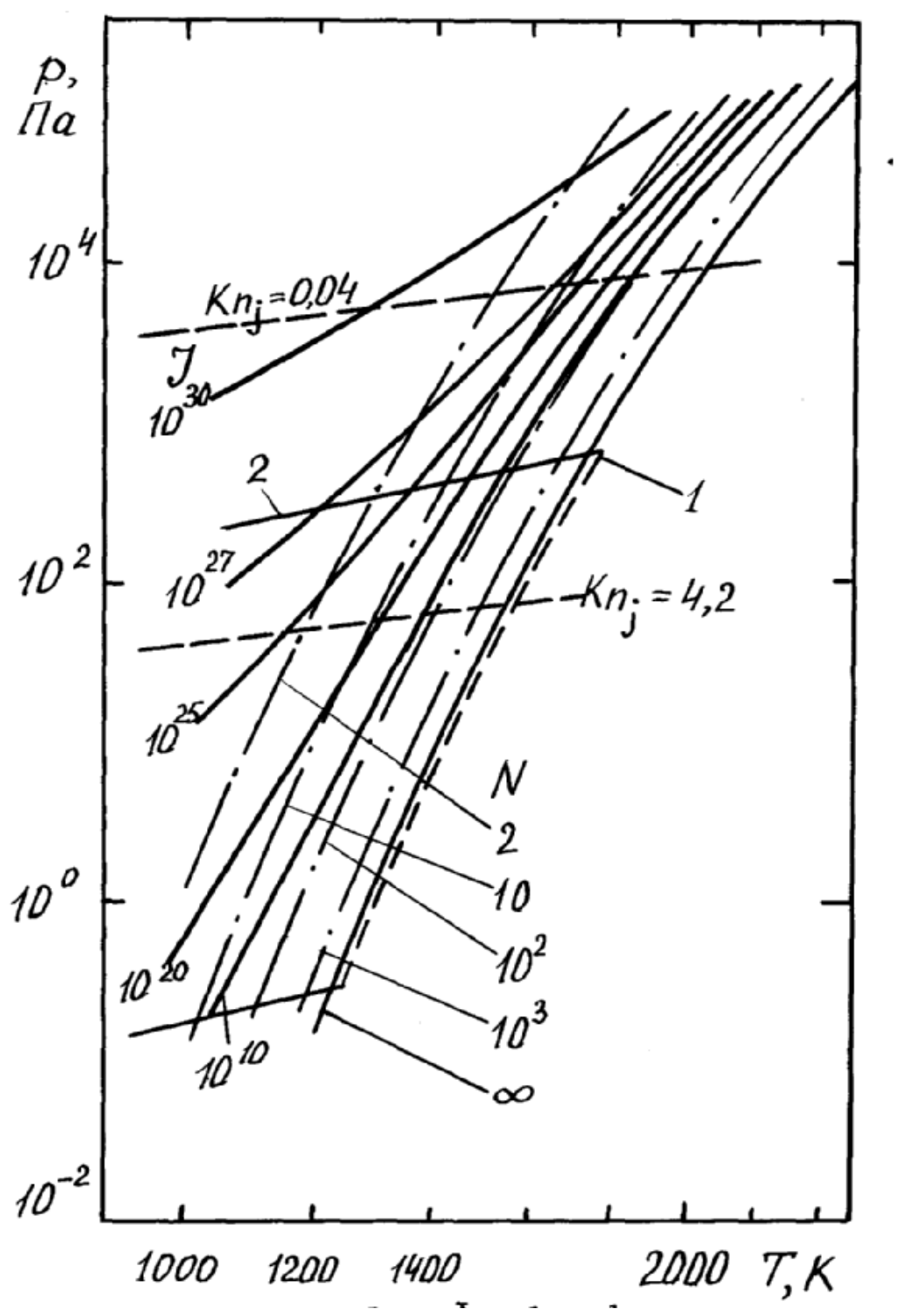

**Рис. 6-1**. Расчетные кривые скорости зародышеобразования и кривые насыщения для кластеров различного размера (3).

Неизотермичность стенок камеры торможения влечёт за собой неизотермичность температурного поля газа в камере. При этом возможно формирование зон пересыщения серебряных паров. В пересыщенном паре может происходить гомогенная конденсация. При гомогенной конденсации сперва, в результате флуктуаций, должен образоваться зародыш жидкой фазы критического размера, который затем может неограниченно расти в размерах. Методика расчета начала гомогенной конденсации паров металлов имеется в (3), соответствующие кривые оттуда приведены на **Рис. 6-1**.

Скорость образования зародышей критического размера рассчитывается по формуле: $J = \left(\frac{P_{Ag}}{kT}\right)^2 V_a \sqrt{\frac{2\gamma_S}{\pi m_{Ag}}} \mathrm{Exp}\left(-\frac{4\pi\gamma_S r_*^2}{3kT}\right)$ . Здесь $P_{Ag}$ – давление паров серебра, $T$ – температура, $V_a$ – объём атома серебра, $m_{Ag}$ – масса атома серебра, $\gamma_S$ – поверхностное натяжение с учётом поправки Талмена: $\gamma_S = \frac{\gamma_\infty}{1+0.6 r_*/r_a}$, $r_*$ – радиус критического зародыша: $r_* = \frac{2\gamma_\infty V_a}{kT \ln {P_{Ag}}/{P_{0Ag}}} - 0.6 r_a$, $P_{0Ag}$ – давление насыщенных паров серебра.

Объём камеры торможения при максимальной загрузке серебра составляет около 380 мм$^3$, или 3.8·10$^{-7}$ м$^3$. При давлении пара 0.5 Па и температуре 880C, пересыщение составляет 9.6 раз. Критический зародыш при таких условиях имеет диаметр 1.35 нм и состоит из 76 атомов серебра. Скорость зародышеобразования составляет 5.4·10$^{-13}$ шт/(м$^3$·сек), или, в объёме реактора, 7.4·10$^{-16}$ шт/час, что, безусловно, является пренебрежимо малой величиной. Формула с поправками (7), связанными с уменьшением коэффициента поверхностного натяжения при уменьшении радиуса капли, в свою очередь, даёт частоту зародышеобразования 2.4·10$^{-7}$ шт/час, что на несколько порядков выше, но, тем не менее, по-прежнему пренебрежимо мало.

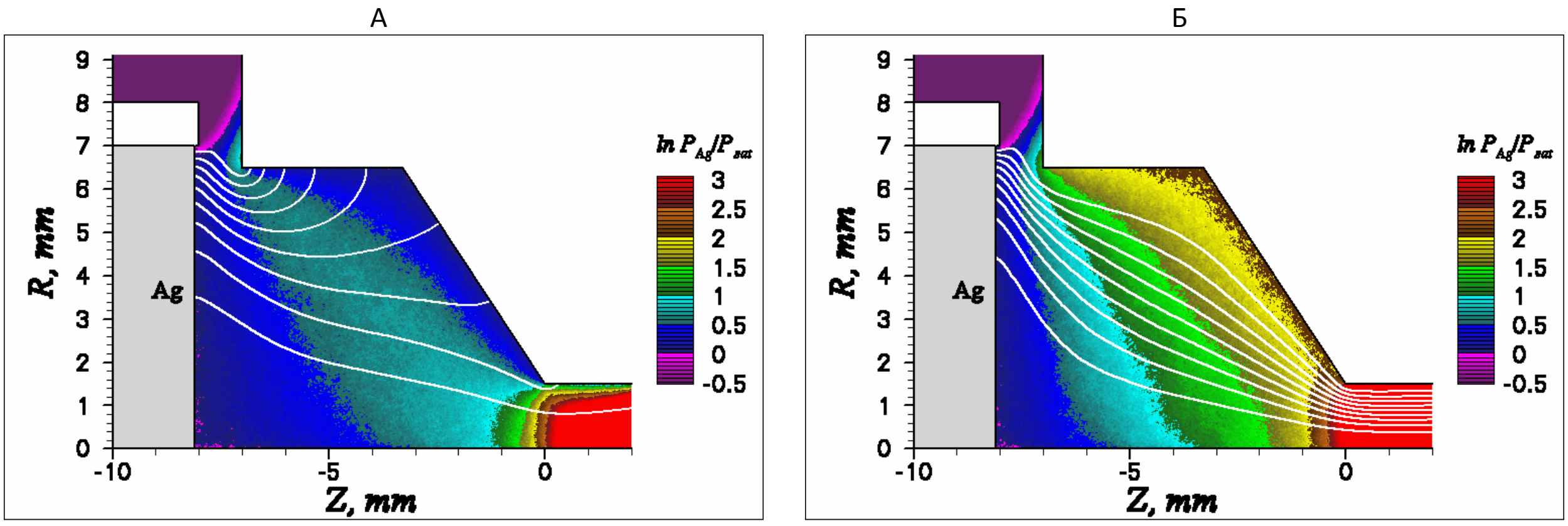

**Рис. 6-2**. Поле пересышения и линии тока серебрянных паров. Слева (А) – полная конденсация и насыщенное испарение на конфузоре (θ = 1), справа (Б) – полное отражение от конфузора (θ = 0).

Как можно было заметить, приведенные оценки являются оценками сверху. На **Рис. 6-2** (а,б) приведены данные ПСМ-расчетов камеры реактора при давлении аргона 3.5 торр: линии тока серебра и рассчитанный по макропараметрам потока натуральный логарифм пересыщения. Результаты приведены для двух крайних коэффициентов конденсации/испарения на холодных поверхностях θ: θ = 1 (а) и θ = 0 (б). В первом случае считалось, что все атомы серебра конденсируются, но в то же время идёт и испарение с потоком, соответствующим давлению насыщенных паров при температуре поверхности. Во втором случае все атомы серебра диффузно отражаются, что в реальности соответствует ситуации, когда поверхность полностью пассивирована неким покрытием («грязью»), препятствующем конденсации серебра.

В обоих случаях, наиболее активно идёт испарение с периферии зеркала расплава, так как там аргон имеет повышенную скорость и «сдувает» серебро, в то время как приосевая часть поверхности находится в «тени». В обоих же случаях во внешней камере реактора (на рисунке показан лишь её фрагмент) наблюдается лишь пренебрежимо малая концентрация паров серебра (при этом считается, что поглощения на её стенках нет), что не удивительно, так как аргон «выдувает» попавшее туда серебро обратно.

В случае (б) линии тока серебра почти всюду неплохо повторяют линии тока аргона, и всё испарённое (и не переконденсировавшееся обратно) серебро выносится в сопло. Вблизи холодной стенки логарифм пересыщения пара превышает два, что согласуется со сделанными выше оценками, но только там и наблюдается такое высокое пересыщение, а не во всей камере реактора, как в сделанных выше оценках. По мере приближения к поверхности расплава пересыщение падает.

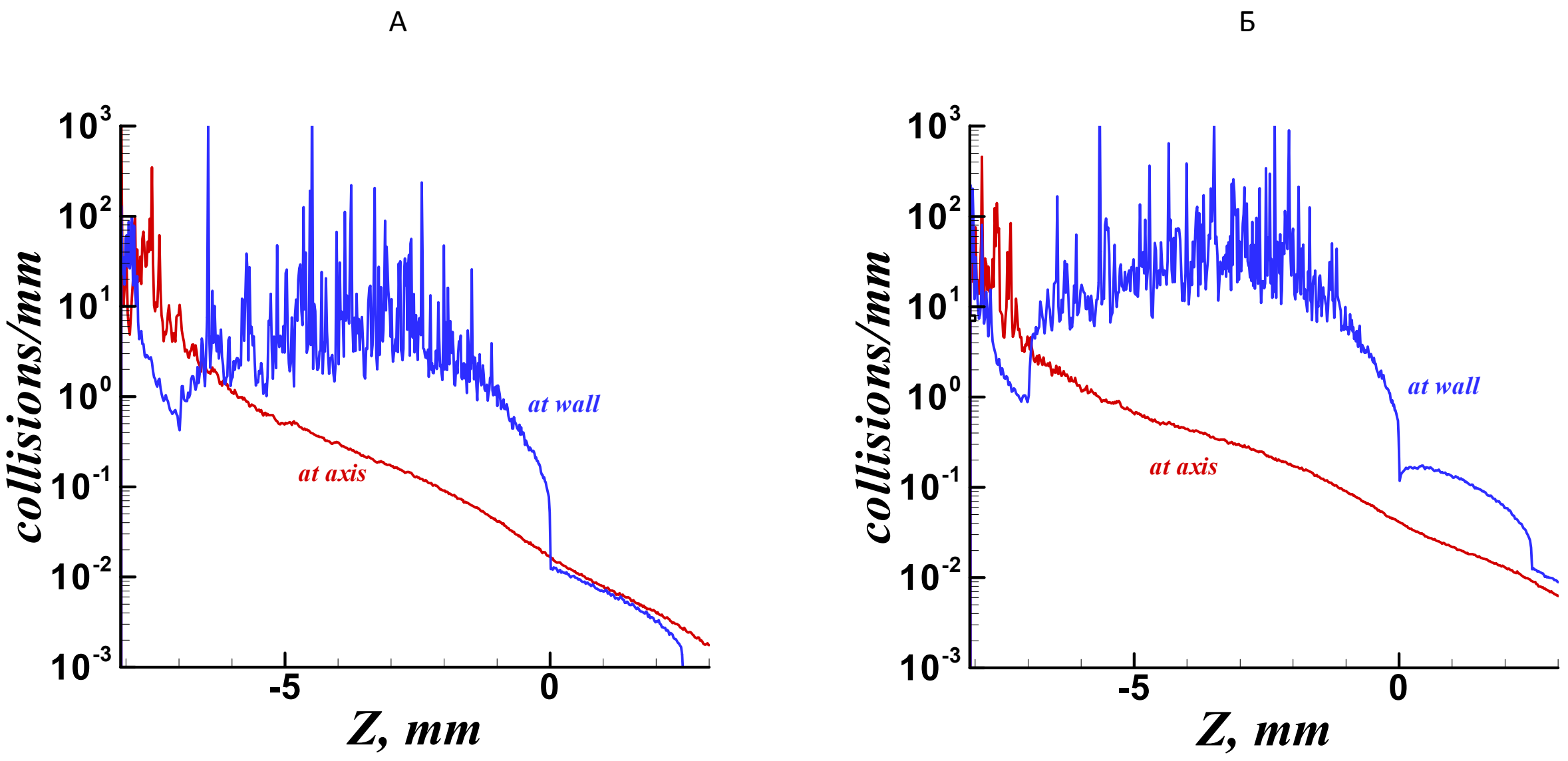

**Рис. 6-3**. Число столкновений на миллиметр смещения вдоль оси.
Красная кривая – на оси, синяя – вдоль стенки конфузора . Слева (А) для случая θ = 1, справа (Б) для θ = 0.

В случае (а) линии тока серебра очень сильно отличаются от линий тока аргона, что согласуется со сделанным при оценках выводом о преимуществе диффузионного переноса серебра над конвективным. На холодных стенках, согласно линиям тока, конденсируется более 80% испаряющегося серебра. Однако, количество испаряющегося серебра при этом выше (вопрос массовых потоков серебра будет рассмотрен далее). Логарифм пересыщения не превышает 1, чему соответствует критический зародыш диаметром 3.5 нм, состоящий из 1300 атомов серебра. Это говорит о том, что в случае больших коэффициентов конденсации гомогенное образование зародышей в реакторе ещё более затруднено.

Дополнительно, имеет смысл проанализировать количество столкновений, которые испытывает атом паров серебра прежде, чем будет вынесен из сопла. Как уже выше отмечалось, между столкновениями серебро-серебро, атом серебра диффундирует в среднем более чем на 2 мм, что не слишком мало по сравнению с диаметром тигля 14 мм, причём диффузия существенна вплоть до числа Маха 0.06, которое реализуется довольно близко от конфузора сопла. Так как разумно предположить, что значительная часть капель может конденсироваться на поверхностях тигля (прежде всего холодных) при столкновении с ними, то активная диффузия затрудняет капле возможность длительное время находиться в зоне максимального пересыщения (т.е. вдали от стенок) и расти, следовательно, ещё более уменьшает скорость зародышеобразования. Следует попробовать, тем не менее, пренебречь диффузией и посчитать количество столкновений серебро-серебро при движении атома вдоль линий тока аргона. Достаточно рассмотреть две линии тока: осевую линию тока и линию тока, проходящую вдоль поверхности тигля (так как течение разрежено, скорость у поверхности не равна нулю).

Впрочем, приведенные оценки сделаны для атомов серебра и потому могут иметь смысл лишь для сравнительно небольших кластеров серебра. Зародыши размером порядка 1 нм и более, если всё же образовались, из-за большого сечения столкновения гораздо хуже диффундируют в аргоне и при этом гораздо чаще сталкиваются с атомами паров серебра. Другими словами, хотя условия для гомогенного зародышеобразования в тигле крайне неблагоприятны, они вполне могут способствовать дальнейшему росту зародышей крупнее критического размера, образовавшихся на поверхности тигля. На **Рис. 6-3** для обоих случаев показаны зависимости среднего количества столкновений серебра с другими атомами серебра за время перемещения на миллиметр вдоль оси. В случае (а), число столкновений на миллиметр для обеих линий тока не превышает 10 всюду, кроме непосредственной близости к поверхности расплава. По мере приближения к критическому сечению сопла (где наблюдается огромное пересыщение пара) число столкновений резко падает до пренебрежимо малых значений. Забегая вперёд, следует сказать, что, даже после торможения сверхзвуковой струи до дозвуковой скорости в фоновом газе, частота столкновений останется пренебрежимо малой. В случае (б), частота столкновений на приосевой линии тока изменилась слабо, и на пристеночной линии тока увеличилась всего в несколько раз. Таким образом, полное количество столкновений серебро-серебро при движении вдоль линий тока не превышает одной-двух сотен. Вновь, это говорит о крайней неблагоприятности условий для гомогенного зародышеобразования в объёме тигля.

# 7. Распределение потока массы испаренного серебра внутри тигля

Нетрудно посчитать, что при температуре зеркала расплава 980°C, чему соответствует давление насыщенных паров 3.54 мторр, и при единичном коэффициенте испарения, испаряется $6.06\cdot10^{-6}$ кг/(м$^2$·с), или 336 мг/час со всей поверхности расплава. Это теоретический предел. Большая часть серебра, однако, сконденсируется обратно.

Рассмотрим сначала случай (б), т.е. при давлении аргона 3.44 торр и нулевом коэффициенте конденсации на холодных стенках тигля. Обратный поток паров серебра в этом случае составляет 93.1 %, и лишь остальные 23.2 мг/час покидают реактор. Это почти на 20 % меньше величины 28.6 мг/час, полученной выше для изотермического реактора. По-видимому, холодный, и из-за этого более плотный и менее вязкий слой газа у стенки способствует «притяжению» линий тока аргона в ущерб горячему слою у поверхности расплава, и, в то же время, ухудшает диффузию серебряных паров вглубь потока аргона.

Ещё более интересен случай (а) единичного коэффициента конденсации на холодных поверхностях. В этом случае, обратный поток уменьшается до 90.1 %, а оставшиеся 33.4 мг/час испаренного серебра распределены следующим образом: 26.4 мг/час осаждается на конфузоре (ещё 71.4 мг/час осаждается, но снова реиспаряется), 2.6 мг/час осаждается на стенках паразитного капилляра длиной 2.5 мм между конфузором и диффузором (ещё 0.4 мг/час реиспаряется). Только оставшиеся 4.4 мг/час (13 %) достигают диффузора. Забегая вперёд, следует сказать, что бóльшая часть паров серебра, попав в диффузор, сконденсируется на его стенках. Сравнение массовых потоков серебра при различных режимах представлено в **Табл. 7.1**.

**Табл. 7.1**. Распределение массовых потоков испаренного серебра для различных режимов.

| Режим / Величина | 0.12 торр, θ = 1 | 0.98 торр, θ = 0 | 0.98 торр, θ = 1 | 3.44 торр, θ = 0 | 3.44 торр, θ = 1 |
|---|---|---|---|---|---|
| Баланс на зеркале расплава | -171.3 мг/час | -17.9 мг/час | -67.1 мг/час | -23.2 мг/час | -33.4 мг/час |
| Обратный поток на зеркало расплава | 49.0 % | 94.7 % | 80.0 % | 93.1 % | 90.1 % |
| Баланс на конфузоре | 167.6 мг/час | 0 | 63.7 мг/час | 0 | 26.4 мг/час |
| Реиспарение на конфузоре | 71.3 мг/час | 653 мг/час | 71.4 мг/час | 560 мг/час | 71.4 мг/час |
| Поверхностное пересыщение на конфузоре | 3.35 | 9.15 | 1.89 | 7.84 | 1.34 |
| Баланс на капилляре | 2.87 мг/час | 0 | 2.31 мг/час | 0 | 2.65 мг/час |
| Реиспарение на капилляре | 0.438 мг/час | 33.2 мг/час | 0.433 мг/час | 26.6 мг/час | 0.442 мг/час |
| Поверхностное пересыщение на капилляре | 7.55 | 76.7 | 6.33 | 60.2 | 7.00 |
| Баланс через сечение начала диффузора | 0.7 мг/час | 16.8 мг/час | 1.2 мг/час | 23.3 мг/час | 4.6 мг/час |

Режим с давлением аргона 0.12 торр соответствует переходному режиму реактора, когда реактор уже прогрелся (или ещё не остыл), но подача аргона отключена. Дело в том, что в экспериментах, в целях экономии газа, поток аргона подавался только во время основной фазы эксперимента, обычно длящейся не более 30 минут. Выход реактора на режим (прогрев) длился порядка 15-20 минут, примерно столько же длилось остывание реактора. Причём около половины этого времени температура в реакторе близка к температуре плавления серебра, при которой давление насыщенных паров всего в 1.5 раза меньше, чем в рабочем режиме. Как видно, суммарная длительность такого переходного режима в начале и конце эксперименте сравнима с длительностью основной фазы эксперимента. Поэтому на него полезно обратить внимание, если интересны процессы, происходящие на поверхностях внутри тигля. Состояние этих поверхностей в переходном режиме может существенно измениться, что, затем, может оказать влияние на поверхностные процессы во время основной фазы эксперимента.

Из таблицы видно, что, со снижением давления в реакторе, уменьшается поток серебра, достигающий диффузора, и увеличивается поток серебра, конденсирующийся на конфузоре. Баланс серебра на паразитном капилляре изменяется слабо. В таблице также для конфузора и капилляра приведены отношения падающего на поверхность потока серебра к потоку, соответствующему давлению насыщенного пара при температуре поверхности, такое отношение названо поверхностным пересыщением. Также приведена величина обратного потока испарившегося серебра на зеркало расплава. Обратный поток – фактически, то же, что и поверхностное пересыщение, но относится к испаряющей, а не конденсирующей поверхности, из-за чего не превышает единицу, и его удобнее измерять в процентах.

# 8. Течение смеси аргон + пары серебра из реактора-испарителя

Выше приводилась сводная **Табл. 7.1** о массовых потоках паров серебра в тигле. Однако, куда больший интерес представляет величина потока паров серебра на подложку-мишень. В экспериментах (2) использовались подложки различного размера и формы. Для определенности, расчеты течения газа снаружи тигля проводились с дискообразной подложкой диаметром 23 мм, установленной в 52 мм от критического сечения сопла. В данных расчетах использовалось несколько меньшие пространственное разрешение и число частиц внутри реактора-испарителя. Кроме того, внутренняя геометрия реактора была упрощена, а также использовалась иная постановка граничных условий. Последнее привело к тому, что давление в реакторе несколько отличается от расчетов предыдущего подраздела. Режим с давлением 2 торр близок к режиму проведенного эксперимента по осаждению серебра на узкую пластину, который будет описан далее.

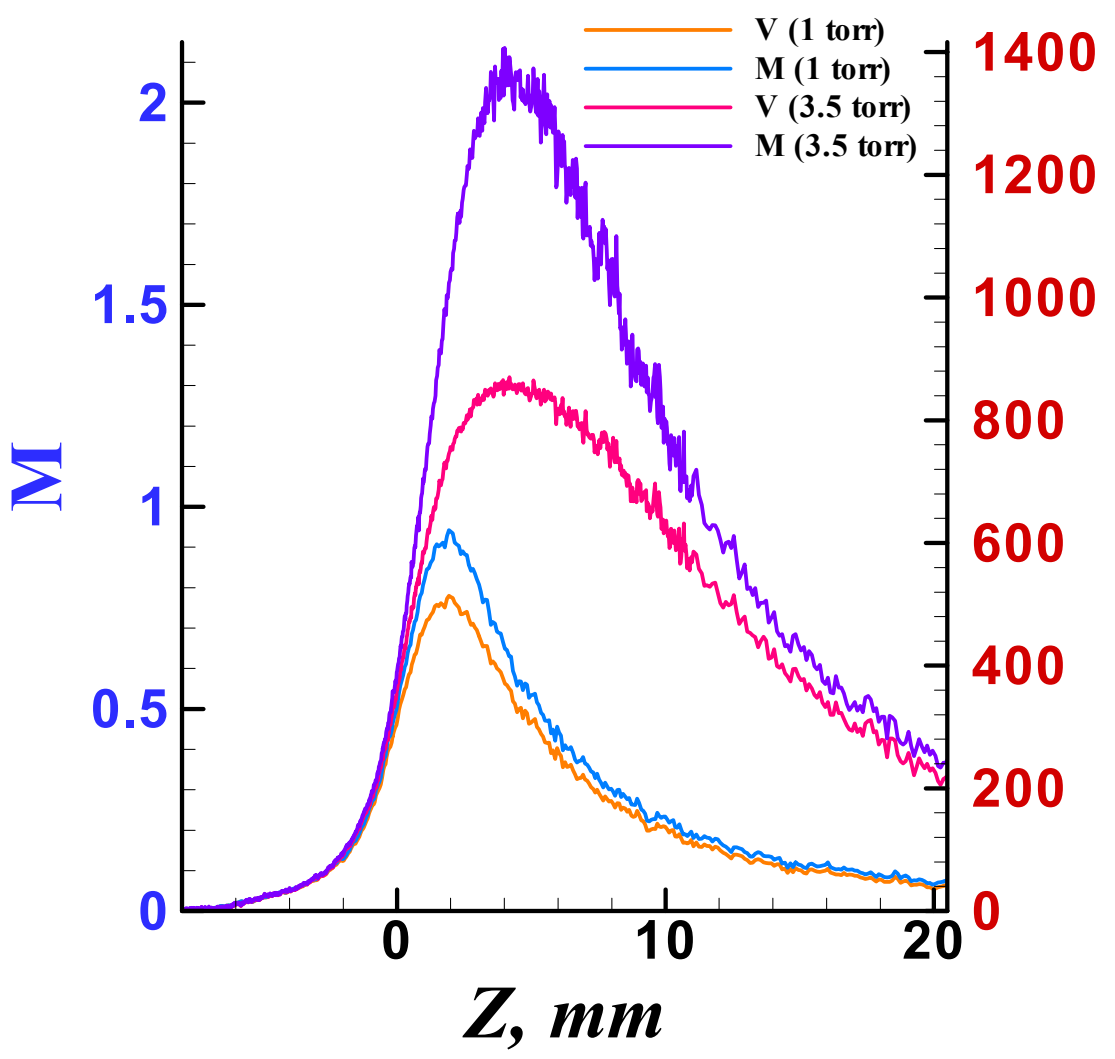

**Рис. 8-1**. Скорость и число Маха аргона у оси струи: внутри тигля и в диффузоре сопла. Сравнение двух режимов.

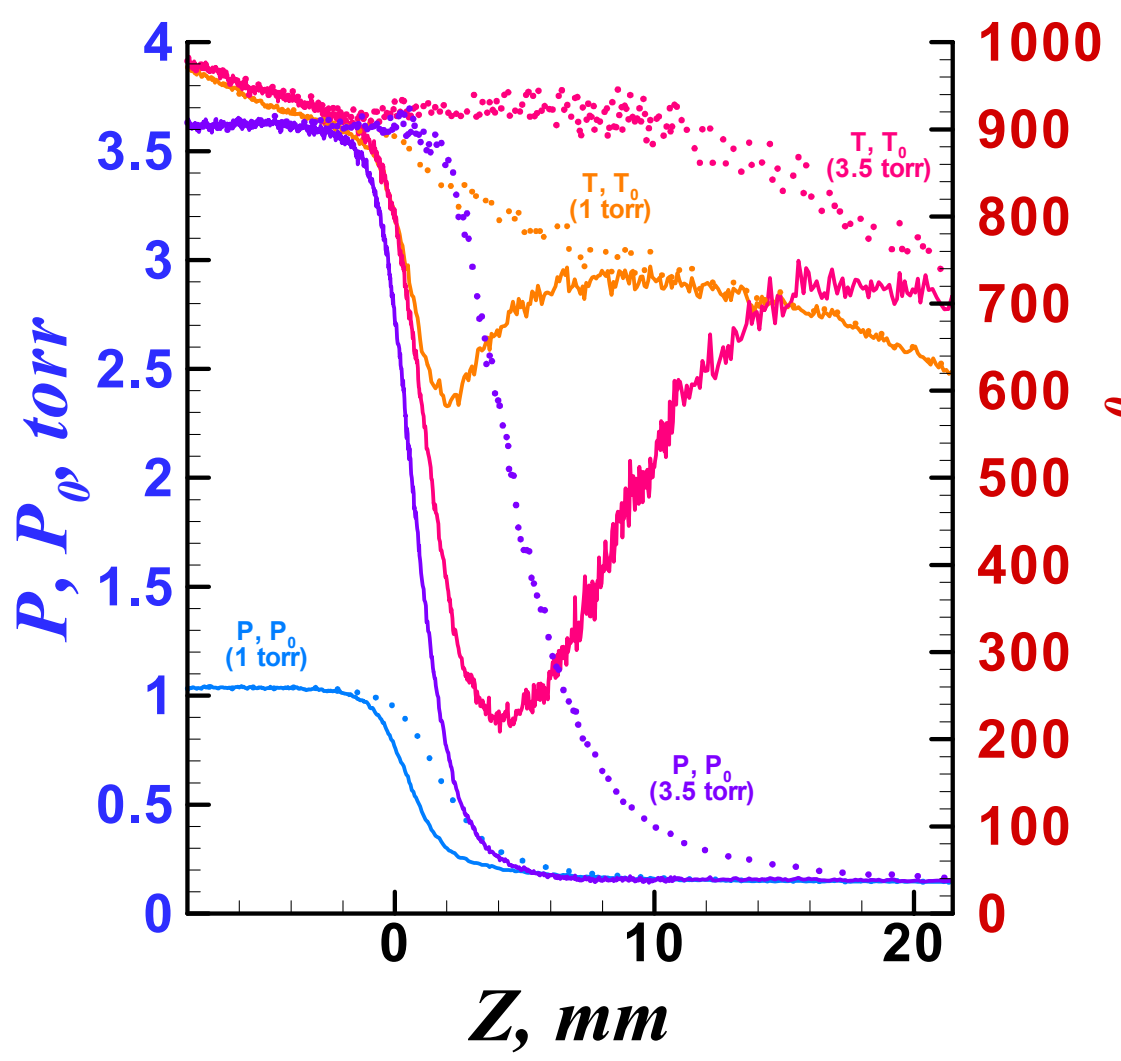

**Рис. 8-2**. Температура, полная энтальпия, давление, полное давление аргона внутри тигля и в диффузоре сопла. Сравнение двух режимов.

На **Рис. 8-1** показано поведение скорости аргона и числа Маха на оси потока вплоть до среза сверхзвукового сопла, для режимов 1 давлением аргона 1 торр и 3.5 торр. До начала капилляра поведение этих параметров практически совпадает. Далее газ начинает стремительно расширяться и ускоряться. Неожиданным оказалось то, что в режиме 1 торр разгон аргона прекращается на числе Маха M≈0.95, т.е. звуковая скорость в сопле так и не достигается, несмотря на более чем 7-кратный перепад давления. В режиме 3.5 торр, число Маха аргона достигает M≈2.05. Достигнув максимальной скорости, газ начинает замедляться. В обоих режимах наблюдается плавное торможение потока, без образования отчётливых ударных волн, что, опять же, весьма неожиданно для режима 3.5 торр. Плавное торможение можно объяснить разреженностью потока – снижение числа Маха с 2 до 1 происходит в области с характерными размерами порядка десятка-двух длин свободного пробега и, кстати, сравнимыми и с диаметром сопла. До среза сопла скорость успевает снизиться до ≈38 м/с в режиме 1 торр и до ≈209 м/с в режиме 3.5 торр, т.е. отличается в 5.5 раз.

На **Рис. 8-2** приведены давления – статическое и полное – для обоих режимов, а также температуры – газа и соответствующая полной энтальпии газа. В обоих режимах наблюдается заметное падение полной энтальпии вследствие охлаждения газа поверхностью сопла. В режиме 1 торр также имеет место стремительное падение полного давления, связанное с потерями на трение в капилляре. Именно этим, совместно со стоком тепла (препятствующему разгону газа при его дозвуковой скорости) и объясняется недостижение скорости звука при сравнительно большом перепаде давления. В режиме 3.5 торр, полное давление слабо меняется вплоть до достижения звуковой скорости, наиболее же резкий спад полного давления приходится на область M > 1.4. В обоих режимах имеет место заметное снижение полной энтальпии газа при течении через сопло – на оси у среза сопла, газ охлаждается до ≈600°C в режиме 1 торр и до ≈750°C в режиме 3.5 торр. Режимы течения при разном давлении сравниваются в сводной **Табл. 8.1**.

На расстоянии 32 мм от среза сопла была установлена поглощающая серебро мишень диаметром 20 мм и с температурой 440°C. При этом определялись скорости осаждения (количество поглощенного серебра в единицу времени), как на лицевой, так и на тыльной сторонах. Предполагая, что осаждается сплошная пленка серебра с плотностью ≈10 г/см$^3$, можно оценить, что при скорости осаждения 1 мкг/(см$^2$·час) толщина пленки растёт со скоростью 1 нм/час.

Как можно видеть, поток серебра на тыльную сторону достигает 40 % от потока серебра на лицевую сторону. Как следствие, в эксперименте следует ожидать осаждения серебра на обе стороны мишени. В режиме с покоящимся аргоном также имеет место доставка серебра на мишень – в основном благодаря диффузионным процессам, но поток в несколько раз меньше.

Забегая вперёд, нужно отметить, сделанные выше выводы не сходятся с результатами экспериментов. Прежде всего, в экспериментах наблюдалась большая скорость осаждения – многие сотни нанометров в час. Расчеты допускают такую скорость осаждения, только если поглощение серебра на поверхностях сопла и тепловых экранах пренебрежимо мало. Во-вторых, в экспериментах скорости осаждения на разные стороны мишени отличались многократно: если на лицевой поверхности возникал хорошо различимый глазом блестящий слой серебра, то на тыльной стороне видимых следов осаждения серебра не было.

**Табл. 8.1**. Распределение массовых потоков испаренного серебра для различных режимов.

| Режим / Величина | 0.14 торр, θ = 1 | 1.04 торр, θ = 0 | 1.04 торр, θ = 1 | 2.02 торр, θ =0 | 2.02 торр, θ = 1 | 3.63 торр, θ = 0 | 3.63 торр, θ = 1 |
|---|---|---|---|---|---|---|---|
| Баланс на зеркале расплава | -164 мг/час (-120 мкм/час) | -23.1 мг/час (-17 мкм/час) | -63.8 мг/час (-46 мкм/час | -27.5 мг/час (-20 мкм/час | -43.2 мг/час (-31 мкм/час | -25.3 мг/час (-18 мкм/час | -31.0 мг/час (-22 мкм/час |
| Обратный поток на зеркало расплава | 43.3 % | 92.0 % | 78.0 % | 90.5 % | 85.1 % | 91.3 % | 89.3 % |
| Баланс на конфузоре | 161 мг/час | 0 | 58.9 мг/час | 0 | 36.5 мг/час | 0 | 21.8 мг/час |
| Реиспарение на конфузоре | 71.4 мг/час | 620 мг/час | 71.3 мг/час | 513 мг/час | 71.5 мг/час | 438 мг/час | 71.4 мг/час |
| Поверхностное пересыщение на конфузоре | 3.25 | 8.68 | 1.83 | 7.17 | 1.51 | 6.13 | 1.31 |
| Баланс на капилляре | 1.70 мг/час (17 мкм\час) | 0 | 1.44 мг/час (15 мкм\час) | 0 | 1.41 мг/час (14 мкм\час) | 0 | 1.28 мг/час (13 мкм\час) |
| Реиспарение на капилляре | 0.17 мг/час | 12.1 мг/час | 0.175 мг/час | 8.72 мг/час | 0.174 мг/час | 7.20 мг/час | 0.178 мг/час |
| Поверхностное пересыщение на капилляре | 10.0 | 69.0 | 9.23 | 50.1 | 9.10 | 40.4 | 8.19 |
| Баланс на диффузоре | 1.47 мг/час (160 нм/час) | 0 | 2.74 мг/час (290 нм/час) | 0 | 4.11 мг/час (440 нм/час) | 0 | 4.53 мг/час (480 нм/час) |
| Реиспарение на диффузоре | 2.17 мг/час | 167.3 мг/час | 2.14 мг/час | 89.9 мг/час | 2.17 мг/час | 48.3 мг/час | 2.12 мг/час |
| Поверхностное пересыщение на диффузоре | 1.7 | 78.0 | 2.28 | 42.4 | 2.89 | 23.8 | 3.14 |
| Баланс через срез сопла | ~0.25 мг/час | 22.9 мг/час | 0.5 мг/час | 25.2 мг/час | 1.1 мг/час | 24.5 мг/час | 3.4 мг/час |
| Реисп.(θ=0)/конд.(θ=1) на стенке-экране | 0.19 мг/час | 111.9 мг/час | 0.388 мг/час | 74.1 мг/час | 0.810 мг/час | 58.3 мг/час | 1.77 мг/час |
| Баланс на лицевой поверхности мишени | ~5 мкг /(см$^2$·час) | 835 мкг /(см$^2$·час) | 9 мкг /(см$^2$·час) | 930 мкг /(см$^2$·час) | 25 мкг /(см$^2$·час) | 920 мкг /(см$^2$·час) | 93 мкг /(см$^2$·час) |
| Баланс на тыльной поверхности мишени | <2 мкг /(см$^2$·час) | 375 мкг /(см$^2$·час) | ~3 мкг /(см$^2$·час) | 400 мкг /(см$^2$·час) | 10 мкг /(см$^2$·час) | 385 мкг /(см$^2$·час) | 37 мкг /(см$^2$·час) |

# 9. Модель расчета движения нанокластеров серебра в потоке аргона

Теперь, уже когда предположение о росте покрытия из нанокластеров и наночастиц серебра можно считать практически доказанным, закономерным этапом работы будет расчет движения нанокластеров серебра в потоке аргона. Результаты расчетов можно затем сравнить с полученным профилем покрытия.

Вообще говоря, при движении кластеров возможно как их испарение, так и дальнейший рост за счёт конденсации на них паров. Выше было показано, что в реакторе наиболее вероятны небольшие пересыщения, при которых критический размер зародыша составляет не менее 3.5 нм, а значит, в реакторе более лёгкие кластеры скорее склонны испаряться. Запасенной в кластерах размером порядка 1 нм энергии заведомо недостаточно для их испарения. Следовательно, для испарения необходим эффективный обмен энергией с аргоном. Так, для кластера из 64 атомов требуется по крайней мере порядка 100 столкновений для релаксации его внутренней энергии с температурой аргона.

Впрочем, кроме процессов испарения-конденсации, есть и другие неизвестные параметры. Например, изначальное распределение по размерам кластеров, сгенерированных в разных точках неизотермической, вообще говоря, поверхности. Также затруднительно подробно описать процесс плавления нанокластеров – как температура, так и теплота плавления могут заметно отличаться от таковых для макроскопических серебряных тел.

**Нанокластер как молекула.** Учитывая описанную выше неопределенность, было принято решение использовать максимально простую модель. А именно: кластеры представлены твёрдыми шарами, и взаимодействие их с аргоном описывается моделью твердых сфер (для кластеров из 4 атомов – VSS) с применением модели Боргнакке-Ларсена для описания релаксации внутренней энергии. Теплоёмкость таких кластеров полагалась постоянной, соответствующей $6N-9$ внутренним степеням свободы, где $N$ – число атомов в кластере. Процессы плавления и испарения не учитывались, как не учитывалось и взаимодействие с парами серебра. Расчеты проводились для кластеров $N=4, 16, 64, 256$. Параметры модели VSS для столкновения кластеров с аргоном были оценены из общих соображений и выбраны следующими:

| $N$ | 4 | 16 | 64 | 256 |
|---|---|---|---|---|
| Диаметр нанокластера | 0.51 нм | 0.81 нм | 1.28 нм | 2.03 нм |
| Столкновительный диаметр | 0.535 нм | 0.566 нм | 0.802 нм | 1.177 нм |
| ω | 0.89 | 0.5 | 0.5 | 0.5 |
| α | 1.45 | 1 | 1 | 1 |

Для больших кластеров использована модель твёрдых сфер, согласно которой ударившийся атом аргона рассеивается в случайном направлении. Можно, например, трактовать это так: атом аргона прилипает к наночастице, которая вращается, и, через некоторое случайное время, покидает её. Оценка периода обращения 256-атомного нанокластера с вращательной температурой 1253 K составляет порядка 0.1 наносекунд, в то время как среднее время между столкновениями наночастицы с аргоном составляет не менее 10 наносекунд. Таким образом, если положить, что время жизни комплекса из атома аргона и нанокластера серебра лежит между 0.1 нс и 10 нс, то считать рассеяние атомов аргона на кластере серебра изотропных вполне оправдано.

Так как сечения столкновения аргон-аргон и аргон-нанокластер отличаются более чем в 10 раз, это же касается и частот столкновений, приходится использовать компонентные временные множители для замедления нанокластеров, вплоть до 10 раз. Без этого, требовалось бы использовать в 10 раз больше частиц аргона, что сделало бы расчёт неподъёмным на персональном компьютере. Также использованы пространственные временные множители (8) – не только для того, чтобы уменьшить число расчетных частиц в вакуумной камере, где длина свободного пробега в разы больше, но и для того, чтобы ускорить установление там стационарного режима. Несмотря на то, что длина свободного пробега снаружи тигля в разы больше, что означает более активную диффузию, характерные размеры течения тоже большие. Как следствие, характерное время установления течения снаружи тигля нисколько не меньше. Вообще, именно длительность диффузионной релаксации потока нанокластеров является лимитирующим фактором при проведении данных расчетов.

Заметим, что если длина свободного пробега аргона в реакторе при давлении 3.5 торр составляет 75 мкм, то длина свободного пробега 256-атомного нанокластера при этих же условиях составляет порядка 300 нм, что лишь в 150 раз превышает диаметр частицы. Длина свободного пробега наночастицы серебра диаметром 20 нм будет составлять уже порядка размера атома, т.е. в 150 раз меньше диаметра частицы, а средняя тепловая скорость таких частиц составляет всего порядка 1 м/с. Отстраняясь от анализа применимости уравнения Больцмана и модели твёрдых сфер для описания движения таких частиц, заметим лишь, что, при настолько малой длине свободного пробега (и очень малой тепловой скорости), использование метода ПСМ однозначно не оправдано из-за колоссальной вычислительной сложности.

Несмотря на очень малую длину свободного пробега нанокластеров, нет необходимости столь же детально дробить сетку – градиенты несущего газа аргона достаточны плавные, а концентрация примеси нанокластеров пренебрежимо мала, поэтому они выступают фактически как пробные частицы и не воздействуют на течение аргона. Проще говоря, размер ячейки практически не влияет на коэффициент диффузии.

Интересно заметить, что в равновесии под действием гравитации плотность наночастиц диаметром 20 нм согласно распределению Больцмана уменьшается в $e$ раз через каждые 40 мм высоты. Однако, достижение равновесия возможно лишь в полностью покоящемся газе и за довольно большие характерные времена – коэффициент диффузии для таких частиц в аргоне при давлении 0.14 торр и температуре 300 К будет составлять порядка $10^{-4}$ м$^2$/с.

**Нанокластер как макроскопическое тело.** Необходимо упомянуть, что существует ещё один метод расчета движения наночастиц в фоновом газе. В этом случае, наночастица рассматривается как твердое сферическое тело, обтекаемое в свободномолекулярном режиме газом с максвелловским распределением скоростей молекул, при этом траектория движения рассчитывается фактически интегрированием силы сопротивления. Формула для вычисления коэффициента сопротивления следующая (9):

$$C_X = \frac{2s^2+1}{\sqrt{\pi}s^3}\exp(-s^2) + \frac{4s^4+4s^2-1}{2s^4}\operatorname{erf}(s) + \frac{2(1-\varepsilon)\sqrt{\pi}}{3s}\sqrt{\frac{T_W}{T_\infty}}, \qquad s = U_\infty\sqrt{\frac{m}{2kT_\infty}}.$$

Здесь $U_\infty$ – скорость сферы относительно газа, $\sqrt{\frac{2kT_\infty}{m}}$ – наиболее вероятная тепловая скорость молекул газа, $\frac{T_W}{T_\infty}$ – отношение температуры наночастицы к температуре газа, $\varepsilon$ – доля зеркальных отражений от поверхности наночастицы (остальные отражения – диффузные). Как видно из формулы, можно для удобства считать $\varepsilon \equiv 0$, введя при этом соответствующую поправку в отношение $\frac{T_W}{T_\infty}$. Случай полностью зеркального отражения при этом будет соответствовать $\frac{T_W}{T_\infty} = 0$. При $\frac{T_W}{T_\infty} = 1$, коэффициент сопротивления при малых скоростях ($s \ll 1$) увеличивается почти в 1.4 раза по сравнению со случаем $\frac{T_W}{T_\infty} = 0$ , при $s \to \infty$ различие постепенно исчезает.

Ещё раз отметим, что представленная выше формула предназначена для описания макроскопического тела. Размер атома аргона не особо мал по сравнению с нанокластером, поэтому при расчете по этой формуле будет правильнее взять такой диаметр сферы, который равен сумме кинетического диаметра атома аргона (0.33 нм) и диаметра нанокластера. Так же следует обратить внимание, что при диффузном отражении полагается, что отраженный атом исходит из той же точки поверхности тела, на которую попал первоначально. Именно из данного допущения следует вид зависимости силы сопротивления от $\frac{T_W}{T_\infty}$. Для наноразмерных тел это допущение, вообще говоря, не верно. Причина этого не только в том, что характерный период теплового вращения нанокластера может быть достаточно мал по сравнению с временем жизни атома аргона на поверхности, но и также с тем, что адсорбированный атом может активно диффундировать по поверхности тела.

Все эти рассуждения приводят к тому, что полагать отражение аргона от нанокластера полностью диффузным будет некорректно, и, скорее всего, именно зеркальное отражение будет физически более уместным. Стоит упомянуть, что в модели твёрдых сфер полагается именно зеркальное отражение сталкивающихся сферических молекул друг от друга.

Интегрирование силы сопротивления. Понятие о коэффициенте сопротивления само по себе не очень удобно для описания движения нанокластеров через слой газа. Гораздо полезнее будет проинтегрировать эту зависимость и найти, как спадает скорость в зависимости от пройденного пути. Для простоты будем считать, что газ покоится, а его плотность и температура всюду постоянны. Запишем уравнение сохранения энергии в дифференциальной форме:

$$\frac{d\frac{MU^2}{2}}{dx} = F = -n\sigma \cdot \frac{mU^2}{2} \cdot C_X(s), \qquad s = \frac{U}{c}, \qquad c = \sqrt{\frac{2kT}{m}}.$$

Здесь $n, T$ – плотность и температура фонового газа соответственно, $m$ – масса молекулы фоновогоо газа, $M$ – масса нанокластера, $\sigma$ – сечение столкновения нанокластера с фоновым газом, $x$ – координата, $U$ – скорость нанокластера, $F$ – сила сопротивления.

Выразим всюду $U$ через $s$:

$$\frac{Mc^2}{2} 2s \frac{ds}{dx} = -n\sigma \cdot \frac{mc^2}{2} s^2 \cdot C_X(s), \qquad \text{или:}$$

$$\frac{dx}{ds} = -\frac{1}{n\sigma} \cdot \frac{M}{m} \cdot \frac{2}{s \cdot C_X(s)}.$$

Последняя форма позволяет рассчитать расстояние, проходимое наночастицей до остановки:

$$D(s_0) = L_0 \cdot \int_0^{s_0} \frac{2}{s \cdot C_X(s)} ds, \qquad L_0 = \frac{1}{n\sigma} \cdot \frac{M}{m}.$$

Здесь $D$ – расстояние, проходимое частицей до полного погашения её скорости, $s_0$ характеризует начальную скорость. $L_0$ – константа размерности длины, зависящая от плотности газа, сечения столкновения и отношения масс. При малых $s_0$, $D(s_0) \approx \frac{s_0}{\frac{8}{3\sqrt{\pi}} + \sqrt{\pi} \cdot \frac{T_W}{T}} + O({s_0}^3)$.

Отобразим полученную зависимость на графике (**Рис. 9-1**). Представленные кривые не только позволяют определить зависимость пройденного пути, но и, фактически, отображают зависимость, по которой спадает их скорость по мере приближения к точке остановки.

Польза этих зависимостей заключается в следующем. Для достаточно тяжёлых наночастиц, их тепловой скоростью по сравнению с газодинамическими скоростями несущего газа можно пренебречь. Это позволяет рассматривать их как макроскопическое тело, движущееся в свободномолекулярном потоке и использовать формулу для коэффициента сопротивления, пренебрегая эффектами диффузии. Однако, как только скорость движения наночастицы относительно газа сравняется с её тепловой скоростью (которая пренебрежимо мала по сравнению с тепловой скоростью несущего газа), её дальнейшее движение будет определяться уже эффектами диффузии. Другими словами, до точки остановки частицу можно рассматривать как макроскопическое тело, но далее – уже нет.

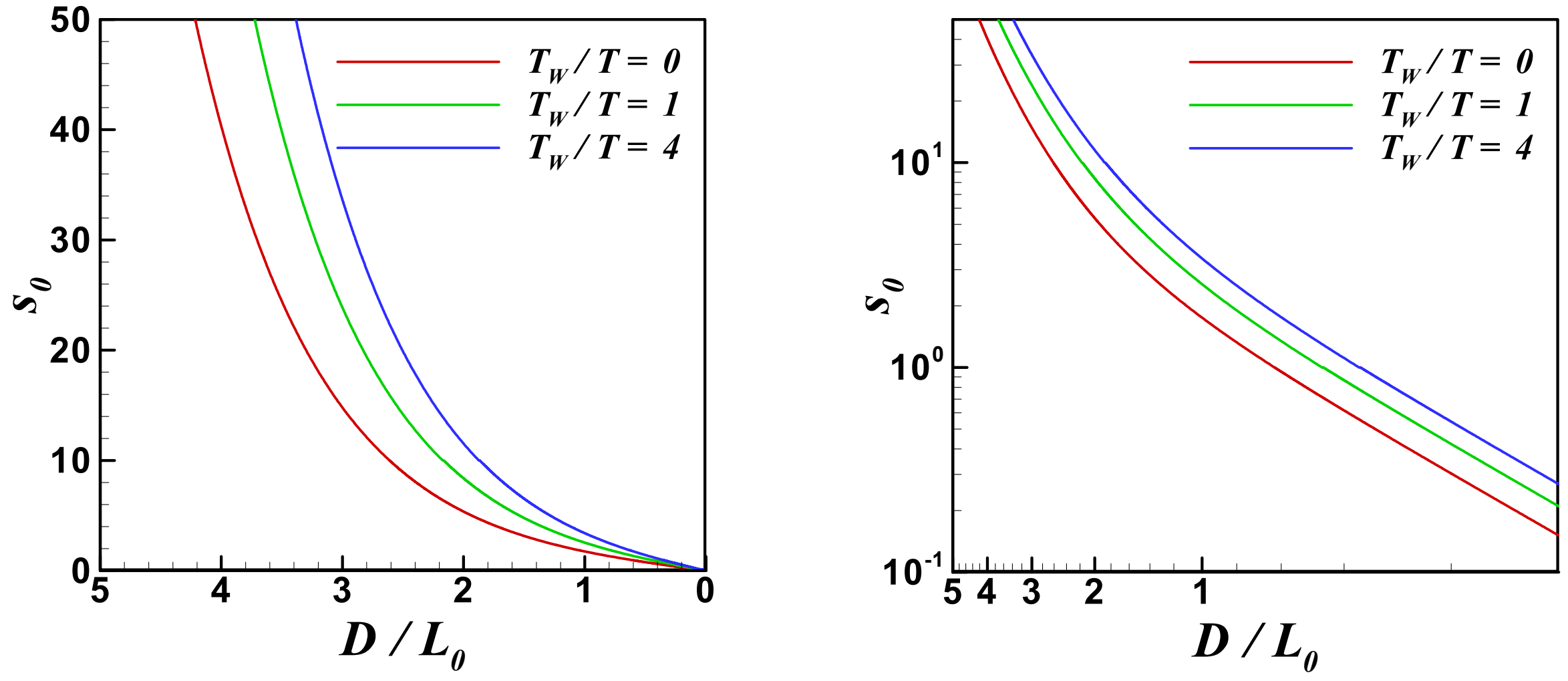

**Рис. 9-1**. Зависимость пути, проходимого наночастицей до потери скорости, в зависимости от начальной скорости, для трёх отношений $\frac{T_W}{T}$. Слева – линейная шкала, справа – логарифмическая шкала.

Сравнение двух моделей. На **Рис. 9-2** показано сравнение скоростей 256-атомных нанокластеров серебра на осевой линии тока, вычисленных двумя способами: ПСМ и интегрированием силы сопротивления. Расчеты выполнены в условиях эксперимента по осаждению серебра на покрытые углеродом медные сетки (диаметром 3 миллиметра, установлены в положении *Z*=32 мм), для двух давлений аргона: 1 и 3.5 торр. При интегрировании использовалось поле скоростей аргона, полученное методом ПСМ, а отражение от частиц полагалось зеркальным. Как видно, результаты интегрирования показывают несколько завышенные по сравнению с ПСМ значения, однако, в целом совпадение весьма неплохое, что свидетельствует о допустимости использования модели твердых сфер для вычисления движения нанокластеров методом ПСМ. Кроме того, в отличие от интегрирования силы сопротивления, использование метода ПСМ позволяет учесть диффузию и термодиффузию.

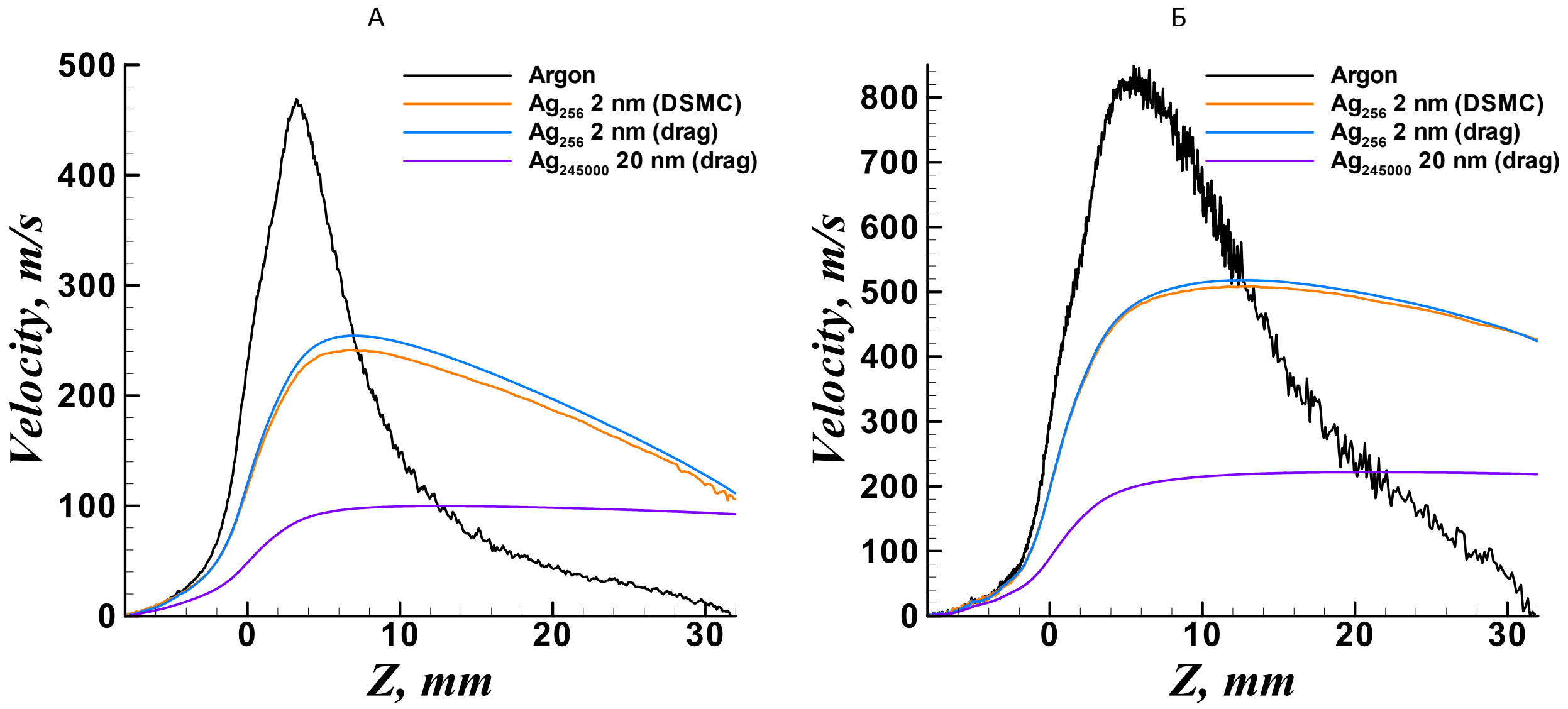

**Рис. 9-2** (А, Б). Скорости на осевой линии тока: аргона (ПСМ), нанокластеров 2 нм (рассчитанные как ПСМ, так и интегрированием силы сопротивления), наночастиц 20 нм (только интегрированием). Режимы с давлением аргона 1 торр (А), и 3.5 торр (Б).

Дополнительно на **Рис. 9-2** показана вычисленная интегрированием силы сопротивления скорость наночастиц диаметром 20 нм.

Как видно, при двух давлениях максимальные скорости наночастиц отличаются примерно в 2 раза (при больших давлениях разгон эффективнее). В обоих случаях к точке *Z*=32 мм нанокластеры (2 нм) теряют порядка 100 м/с своей скорости от максимальной. Однако, в режиме 3.5 торр из-за большой максимальной скорости эта потеря кажется пренебрежимо малой, в то время как в режиме 1 торр теряется порядка ¾ достигнутой кинетической энергии.

Интересно отметить, что, при столкновении с мишенью в виде покрытой углеродом медной сетки, установленной в точке *Z*=32 мм, запас кинетической энергии нанокластеров, если его перевести во внутреннюю энергию, достаточен для их нагрева на 400 К при высоком давлении аргона и всего на 25 К при низком. Такая разница в кинетической энергии, несомненно, может повлиять на процесс взаимодействия нанокластеров с подложкой.

Что касается наночастиц диаметром 20 нм, то они, после разгона, тормозятся очень медленно, сохраняя направленное движение. Впрочем, даже при скорости 200 м/с запас их кинетической энергии небольшой – порядка 85 К.

# 10. Сравнение газодинамики течения при использовании звукового сопла и сверхзвукового сопла

В числе прочего, проводились эксперименты по осаждению серебра на покрытую углеродом медную сетку (2). Эксперимент проводился при двух давлениях аргона (1.5 мм рт. ст. и 3.5 мм рт. ст.), и при двух типах сопла: сверхзвуковое сопло (описанное ранее) и звуковое сопло, исключающее диффузор и паразитный капилляр. Диаметр критического сечения и геометрия конфузора были одинаковы. Мишень диаметром 3 мм устанавливалась в 32 мм от критического сечения. При обоих давлениях результаты эксперимента были следующие: при звуковом сопле осадились только мелкие наночастицы (4-6 нм), а при сверхзвуковом – ещё и крупные (16-20 нм). Из этого был сделан вывод, что крупные наночастицы формируются на поверхности диффузора. Впрочем, новые, совсем недавние эксперименты, обнаружили осаждение крупных наночастиц и из звукового сопла. Несомненно, иная геометрия сопла могла изменить его тепловой режим, что в экспериментах не отслеживалось. Кроме того, несомненно, отсутствие диффузора, способного действовать как поглотитель серебряных паров, могло повысить концентрацию паров серебра вокруг мишени.

Однако, в данном параграфе речь будет идти исключительно о сравнении газодинамики этих двух режимов. Для сравнения был выбран режим с давлением аргона 3.5 мм рт. ст. В расчетах использовалась упрощенная геометрия реактора: смесь аргон + примеси нанокластеров подаётся плоским потоком через границу, соответствующую поверхности расплава, а поверхности сопла поглощают примеси.

На **Рис. 10-1** показаны поля чисел Маха и линии тока для двух режимов. При звуковом сопле, достигаются более высокие числа Маха. Сверхзвуковое сопло оказывает на струю дополнительное тормозящее воздействие, кроме того, паразитный капилляр снижает коэффициент расхода.

Однако, сравнение **Рис. 10-2**, где показаны поля плотности и линии тока нанокластеров массой 64 и 256 атомов, не обнаруживает существенной разницы. Линии тока практически повторяют друг друга, а меньшие абсолютные значения плотности при наличии сверхзвукового сопла объясняются тем, что из-за меньшего коэффициента расхода и, как следствие, уменьшенной скорости течения в конфузоре, примеси более активно поглощаются на стенках.

Как видно, несмотря на заметное отличие параметров струи несущего газа, достаточно тяжелые нанокластеры ведут себя в этих двух струях практически одинаково. Другими словами, оба сопла формируют практически неразличимые потоки нанокластеров.

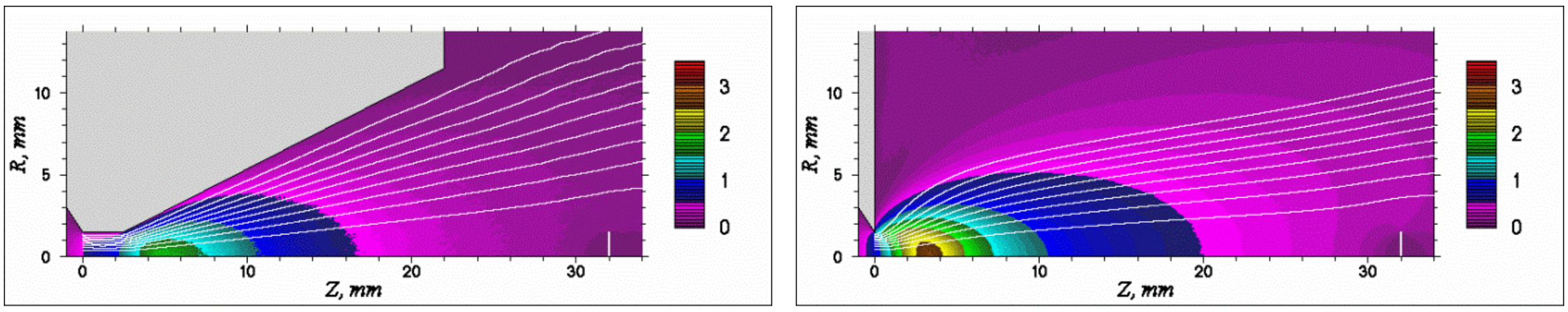

**Рис. 10-1.** Поле чисел Маха и линии тока аргона для сверхзвукового и звукового сопла.

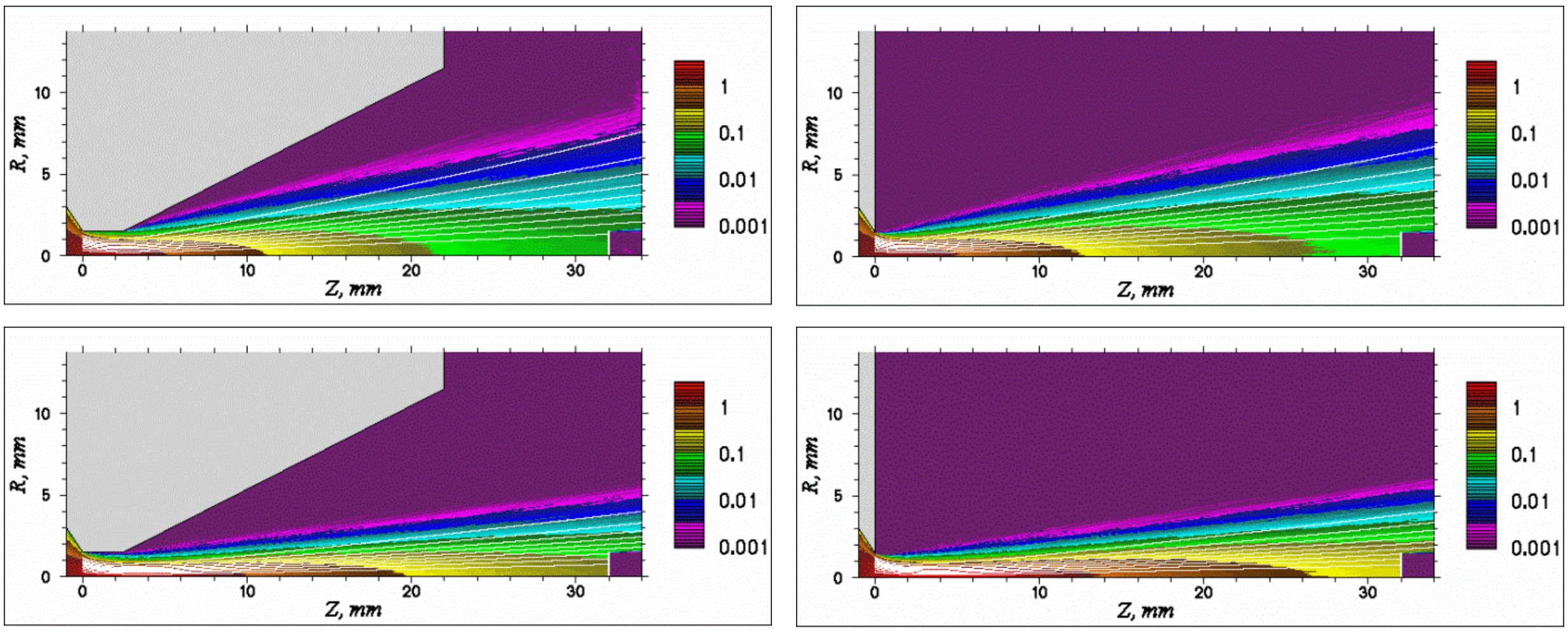

**Рис. 10-2**. Поля плотности и линии тока нанокластеров 64 атома (сверху) и 256 атомов (снизу), для режимов со сверхзвуковым и звуковым соплами.

# 11. Обоснование и результаты эксперимента по осаждению серебра на длинную пластину

Полученное выше противоречие вынуждает задуматься над механизмом доставки серебра на мишень. Прежде всего, разумно предположить, что покрытие серебра формируется из довольно крупных и тяжелых частиц (нанокластеров) серебра – неспособных достаточно оперативно обогнуть мишень, следуя за потоком аргона. Эти частицы не могли образоваться в струе газа, следующего на мишень – концентрация паров серебра крайне низка уже в диффузоре сопла и продолжает стремительно падать при приближении к мишени.

Если предположить формирование кластеров на поверхностях тепловых экранов либо в фоновом газе, то можно заключить следующее. Во-первых, не вполне ясно, почему кластеры проникают через экранирующую струю аргона только на лицевую сторону мишени. Во-вторых, в этом случае профиль пятна на мишени, образованный при осаждении кластеров, как минимум должен быть большим по сравнению с характерным размером струи аргона – т.е. по крайней мере больше диаметра среза сопла. Кроме того, вблизи центра мишени в этом случае скорее возможен провал скорости осаждения, чем максимум.

Остаётся ещё вариант гетерогенного формирования кластеров внутри тигля или сопла. Этот вариант косвенно подтверждается экспериментами по осаждению на покрытую углеродом медную сетку, на которой были обнаружены в том числе и частицы диаметром до 5 нм – причём в экспериментах как со сверхзвуковым соплом, так и со звуковым.

Итак, методом исключения можно прийти к выводу, что наиболее вероятен следующий сценарий. На сравнительно холодных внутренних поверхностях тигля формируются нанокластеры серебра, которые затем захватываются потоком аргона, доставляются на мишень и осаждаются на её лицевой стороне. В плотной трансзвуковой части струи аргона кластеры приобретают избыточную скорость, которая, погашаясь не сразу, некоторое время препятствует диффузионному разлёту кластеров и потому сужает телесный угол их потока и уменьшает их осаждение на стенках диффузора. Имея большой диаметр и, следовательно, меньший в несколько раз коэффициент диффузии в аргоне, эти кластеры значительно хуже огибают мишень.

Были проведены также расчёты движения кластеров серебра различного размера в потоке аргона. Результаты расчетов согласуются со сделанными выводами и будут представлены далее.

Таким образом, был обоснован эксперимент по определению профиля струи кластеров. Было предложено установить в качестве мишени узкую длинную пластину.

В качестве мишени в эксперименте использовалась пластина из нержавеющей стали шириной около 1 см. Пластина была предварительно обработана спиртом и ацетоном. Мишень была установлена на расстоянии 32 мм от среза сопла.

Условия в реакторе при проведении эксперимента были следующими: давление 2.1 торр, температура 990°C. В вакуумной камере давление составляло 20 делений, что при пересчете даёт 0.14 торр. Основная фаза напыления производилась 10 Минут.

На **Рис. 11-1** представлена фотография образца после эксперимента. В середине образца наблюдается ярко выраженный светлый след осажденного серебра длиной 15-20 мм. Также заметны некоторые границы на поверхности, выраженные в резком изменении яркости, в положениях около 20 мм и 60 мм, между ними соответственно около 40 мм. Последние, впрочем, могут быть связаны и с градиентами температуры по длине образца. Обратная сторона образца не имела заметных глазу изменений, поэтому не фотографировалась.

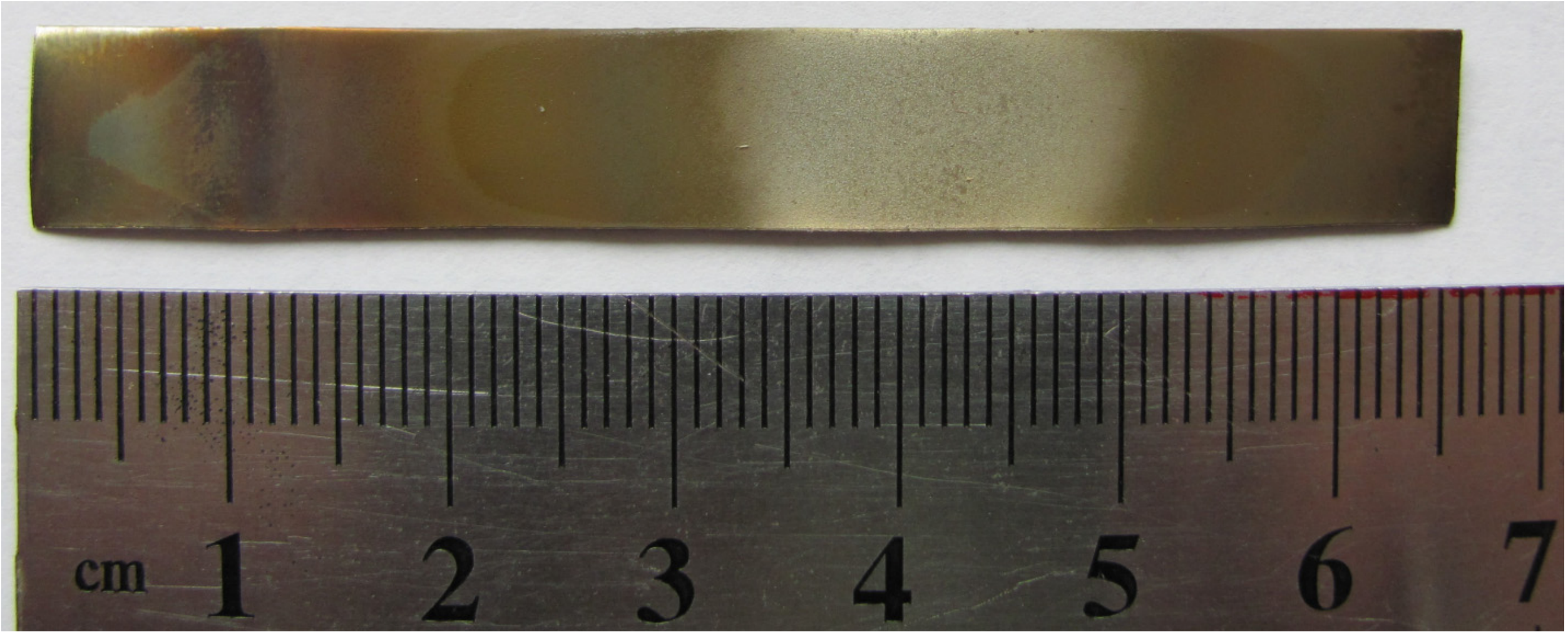

**Рис. 11-1**. Фотография лицевой стороны напыленного образца.

# 12. Движение нанокластеров в потоке аргона при условиях эксперимента

Изучив движения моноатомных паров серебра, начнём для сравнения исследовать движение нанокластеров с самого большого размера – 256 атомов. На **Рис. 9-2** уже демонстрировалось характерное поведение скорости нанокластеров на оси потока, показавшее сильную неравновесность.

На **Рис. 12-1** показано поведение плотностей и скоростей аргона и нанокластеров на оси при условиях эксперимента (расчёт произведен без мишени). Фактически, показаны результаты четырёх вариаций расчета. Во-первых, нанокластеры эмитировались на двух поверхностях: в одном случае – на стенках конфузора (noz), во втором – на поверхности расплава (bot). Во-вторых, рассматривались случаи полного поглощения нанокластеров на любой поверхности (no refl) и полного диффузного отражения от всех поверхностей (100 % refl), кроме той, с которой они эмитированы (на ней они всегда поглощались). Впрочем, для 256-атомных нанокластеров результаты для четырёх случаев отличаются незначительно.

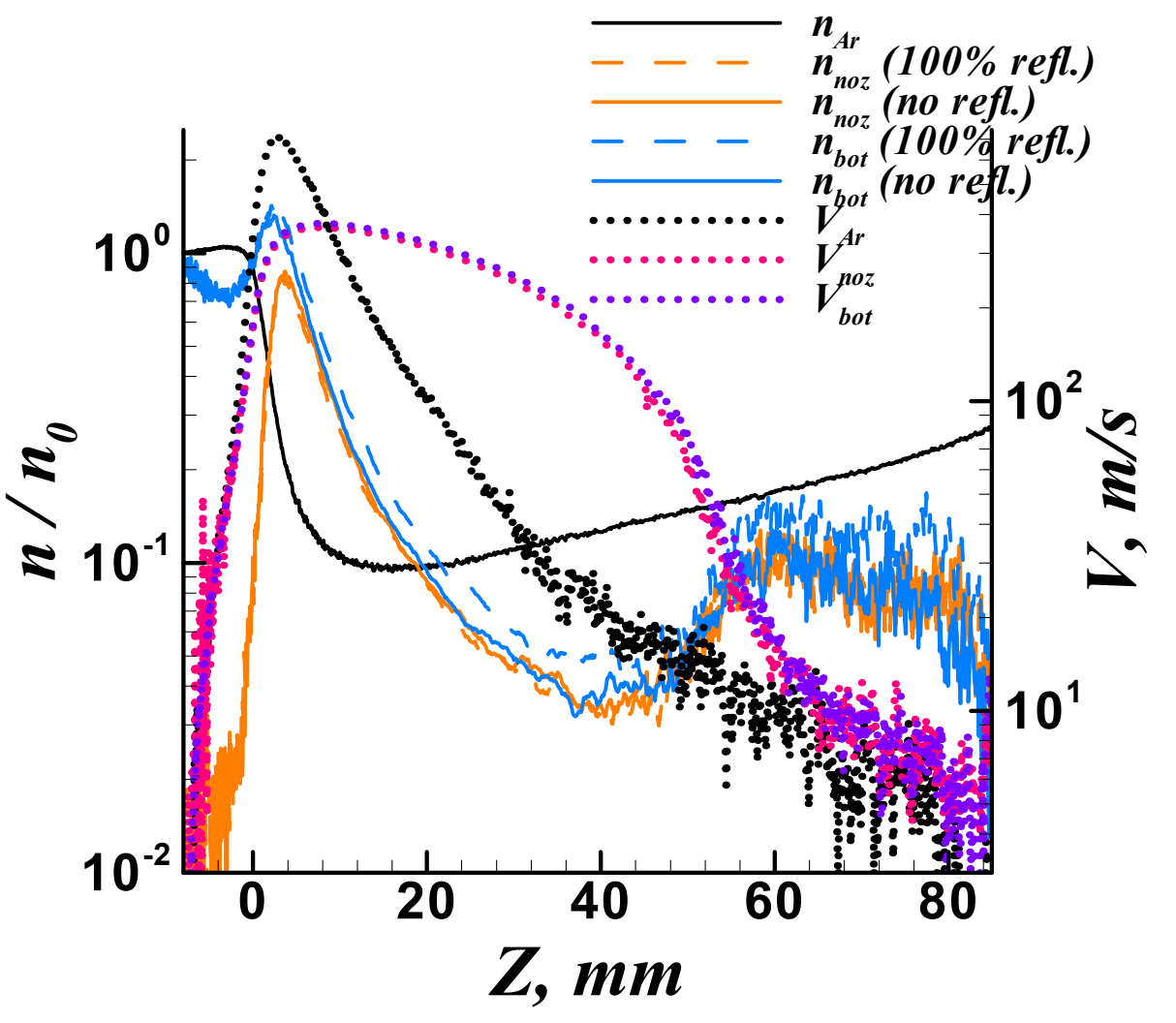

**Рис. 12-1**. Распределение плотности и скорости аргона и нанокластеров (256 атомов) вдоль оси струи.

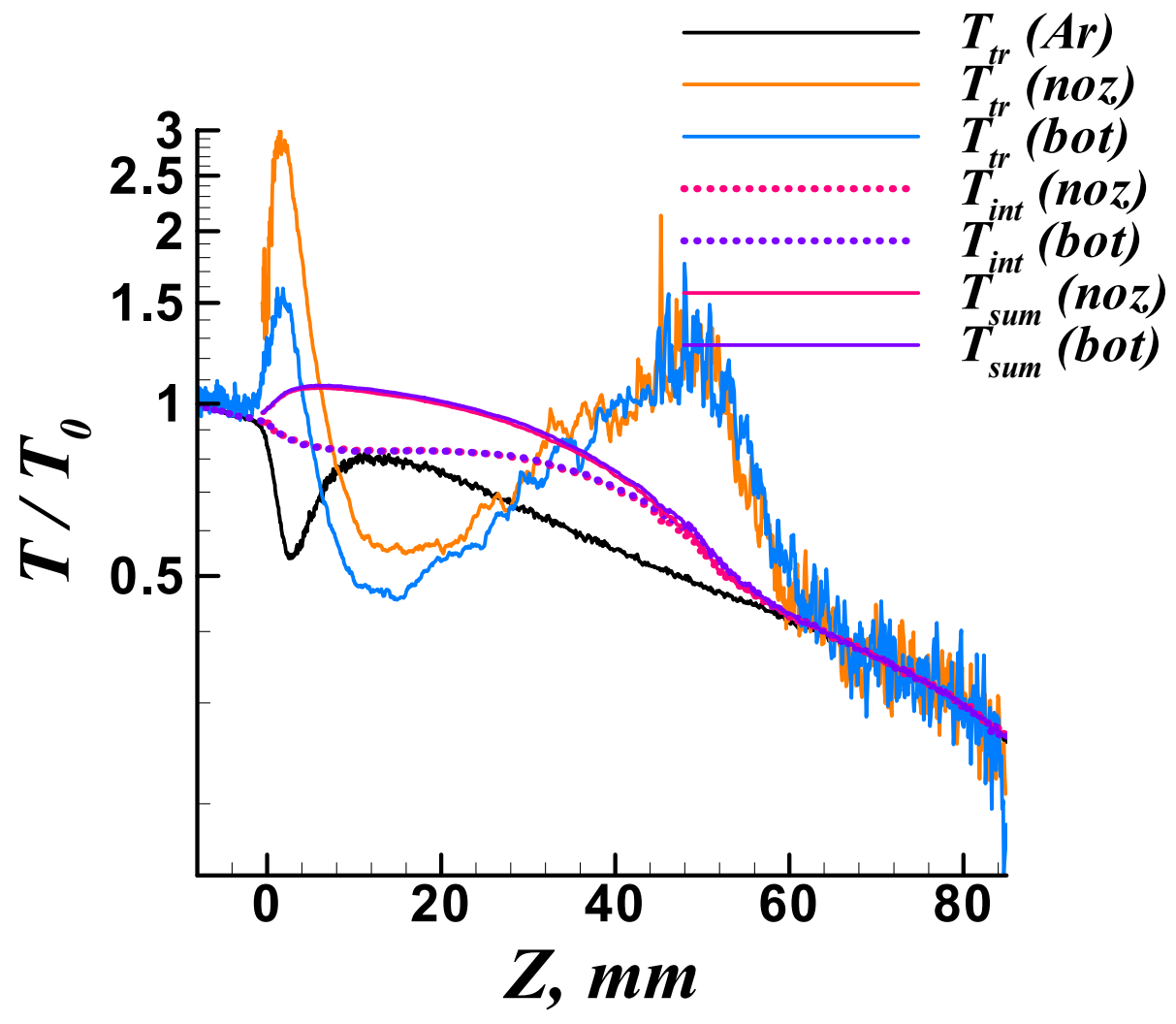

**Рис. 12-2**. Распределение температур аргона и нанокластеров (256 атомов) вдоль оси струи

Единичная плотность $n_0$ имеет следующий смысл. Для аргона она соответствует средней его плотности у поверхности расплава. Для нанокластеров – плотности при «давлении насыщенных паров» и температуре у поверхности расплава. «Давление насыщенных паров» нанокластеров задается одинаковым на всех эмитирующих поверхностях и считается бесконечно малым по сравнению с давлением аргона, эмиссия нанокластеров моделируется как процесс испарения.

Незначительное отличие случаев с поглощением и без него объясняется тем, что нанокластеры почти не соприкасаются со стенками тигля и сопла, хоть и имеют большую вероятность столкнуться с той поверхностью, откуда эмитированы. Наибольшее отличие наблюдается, если кластеры эмитируются поверхностью расплава – в этом случае некоторая доля кластеров достигает поверхности конфузора (и либо поглощается на ней, либо рассеивается). Отсутствие разницы для случая, когда кластеры эмитируются на поверхности конфузора, связано с тем, что при этом кластеры поглощаются на эмитирующей поверхности – конфузоре – уже в обоих вариантах. Влияние же остальных поверхностей – капилляра, диффузора, теплового экрана – незначительно во всех случаях.

Следует ещё раз обратить внимание, что снаружи тигля макропараметры нанокластеров практически совпадают, несмотря на то, что кластеры эмитируются с противоположных поверхностей тигля: в одном случае – с поверхности расплава, в другом – с поверхности конфузора. Следовательно, можно считать, что где именно внутри тигля образовались нанокластеры – неважно с точки зрения параметров сформированной струи.

Снаружи тигля поле макропараметров нанокластеров можно условно разделить на две области. Внутри сопла и снаружи недалеко от него расположена сильно неравновесная область, где имеет место гашение избыточной скорости направленной струи нанокластеров до величины порядка их тепловой скорости. За пределами этой области поток нанокластеров теряет направленность, и их движение становится похожим на равновесную диффузию в аргоне. В данном случае граница лежит в интервале $Z = 50 - 60$ и характеризуется резким подъёмом плотности кластеров и сравнительно быстрым гашением избытка скорости. Как раз примерно там, где в эксперименте установлена мишень ($Z = 52$).

Необходимо оговорить, что, вообще говоря, диффузионная область в представленном расчете несколько искажена, как под влиянием граничных условий, так и тем, что в полной мере достижение её стационарности очень трудоёмко из-за довольно низкого коэффициента диффузии таких нанокластеров ($\sim 10^{-2}$ м$^2$/с, при среднем интервале между столкновениями $\sim$150 нс).

На **Рис. 12-2** представлено поведение различных температур: поступательной аргона, поступательной кластеров, внутренней кластеров $T_{int}$, а также температуры, соответствующей сумме внутренней и кинетической энергии $T_{sum}$ (такую температуру приобретёт кластер, если вся кинетическая энергия перейдёт во внутреннюю). Что касается поступательной температуры кластеров – почти всюду она «живёт своей жизнью», что, вновь, свидетельствует об очень сильной неравновесности. Лишь в области $Z > 60$ она релаксирует к температуре аргона и можно говорить об околоравновесных условиях. Внутренняя температура кластеров быстро снижается до 765°C, после чего вплоть до $Z \approx 35$ держится недалеко от этого значения. Суммарная энергия на срезе сопла практически не отличается от температуры торможения. К месту установки мишени $Z = 52$ внутренняя температура кластеров падает до 365°C, что лишь на 50°C превышает температуру аргона.

Впрочем, для более тяжелых нанокластеров можно ожидать и более высоких значений внутренней и кинетической энергии на подлёте к подложке. Высокие значения энергии кластеров при столкновении с подложкой, в свою очередь, могут сказаться на характере взаимодействия с подложкой. Так, при напылении бактерицидного композита из серебро-политетрафторэтилен, высокие температуры могут привести к химическим реакциям с образованием соединений серебра, фтора и углерода, что может влиять на бактерицидность.

С точки зрения эффективности доставки нанокластеров, мишень, очевидно, следует ставить в неравновесную область, чтобы избежать диффузионного рассеивания струи нанокластеров. Если же поставить мишень в околоравновесную область, то утратившие направленное движение нанокластеры будут частично унесены прочь огибающим мишень потоком аргона. Собственно, в таком режиме как на лицевую, так и на тыльную стороны мишени доставка нанокластеров будет иметь в основном диффузионную природу. При этом потоки на лицевую и тыльную стороны могут быть сопоставимы.

На **Рис. 12-3** – **Рис. 12-8** представлены аналогичные результаты для более лёгких нанокластеров. В целом, тенденция такова, что равновесный режим течения нанокластеров устанавливается всё раньше. Кроме того, обнаруживают себя две закономерности.

Во-первых, чем легче нанокластер, тем сильнее заметна разница плотности для режимов с поглощением нанокластеров на стенках и без поглощения. Это связано с тем, что механизм доставки нанокластеров к поглощающим поверхностям преимущественно диффузионный. Диффузия более эффективна для легких нанокластеров. Кроме того, для тяжелых нанокластеров диффузионный режим устанавливается уже достаточно далеко от этих поверхностей, поэтому достичь их становится сложнее.

Вторая закономерность состоит в том, что для более легких наночастиц кривые плотности для случаев образования нанокластеров на поверхности расплава и на конфузоре сопла расходятся всё сильнее, но только если имеются поглощающие поверхности. При полном отражении нанокластеров, кривые по-прежнему совпадают. Основной вклад в различие плотности вносит поглощающее действие конфузора.

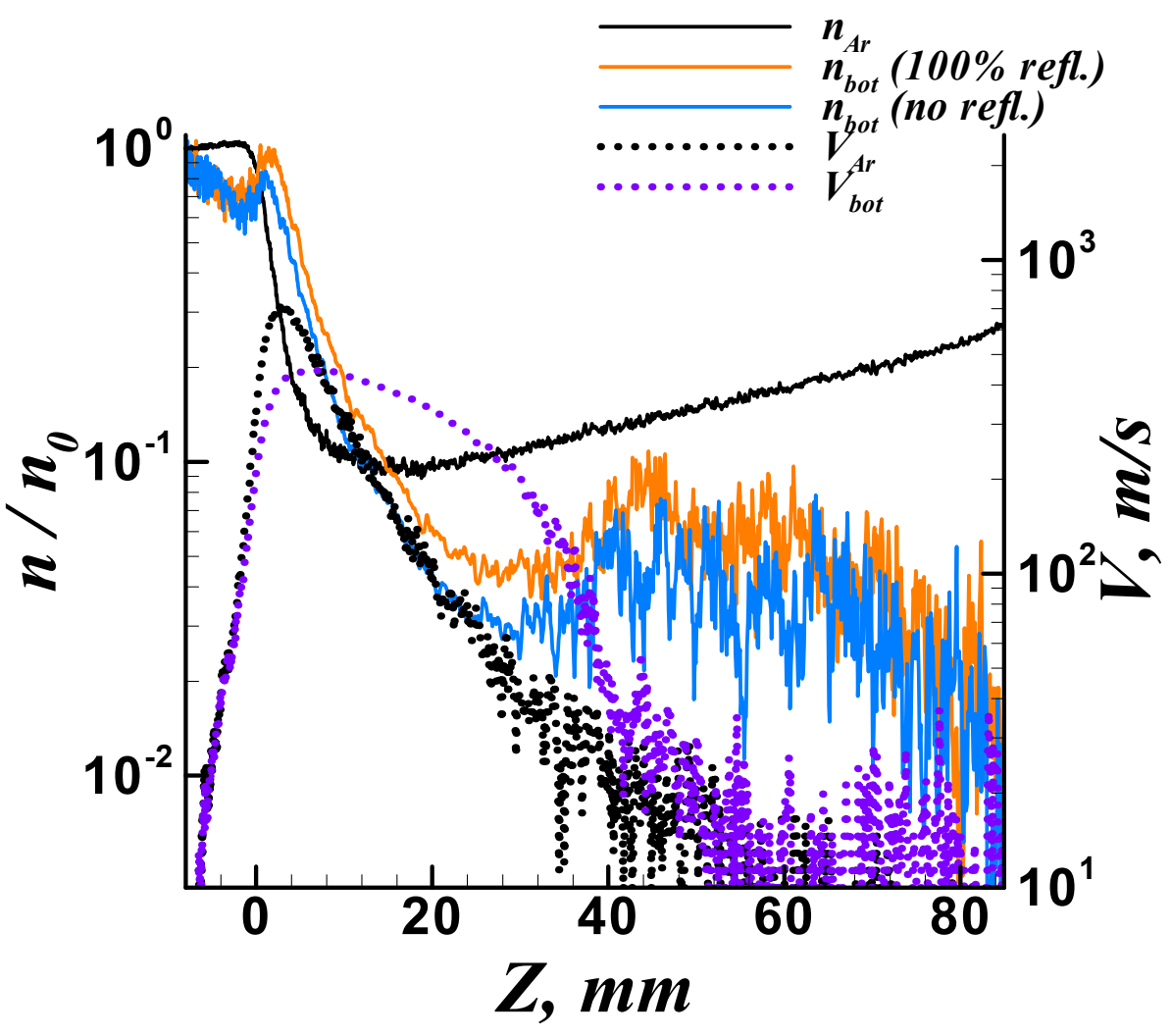

**Рис. 12-3**. Распределение плотности и скорости аргона и нанокластеров (64 атома) вдоль оси струи.

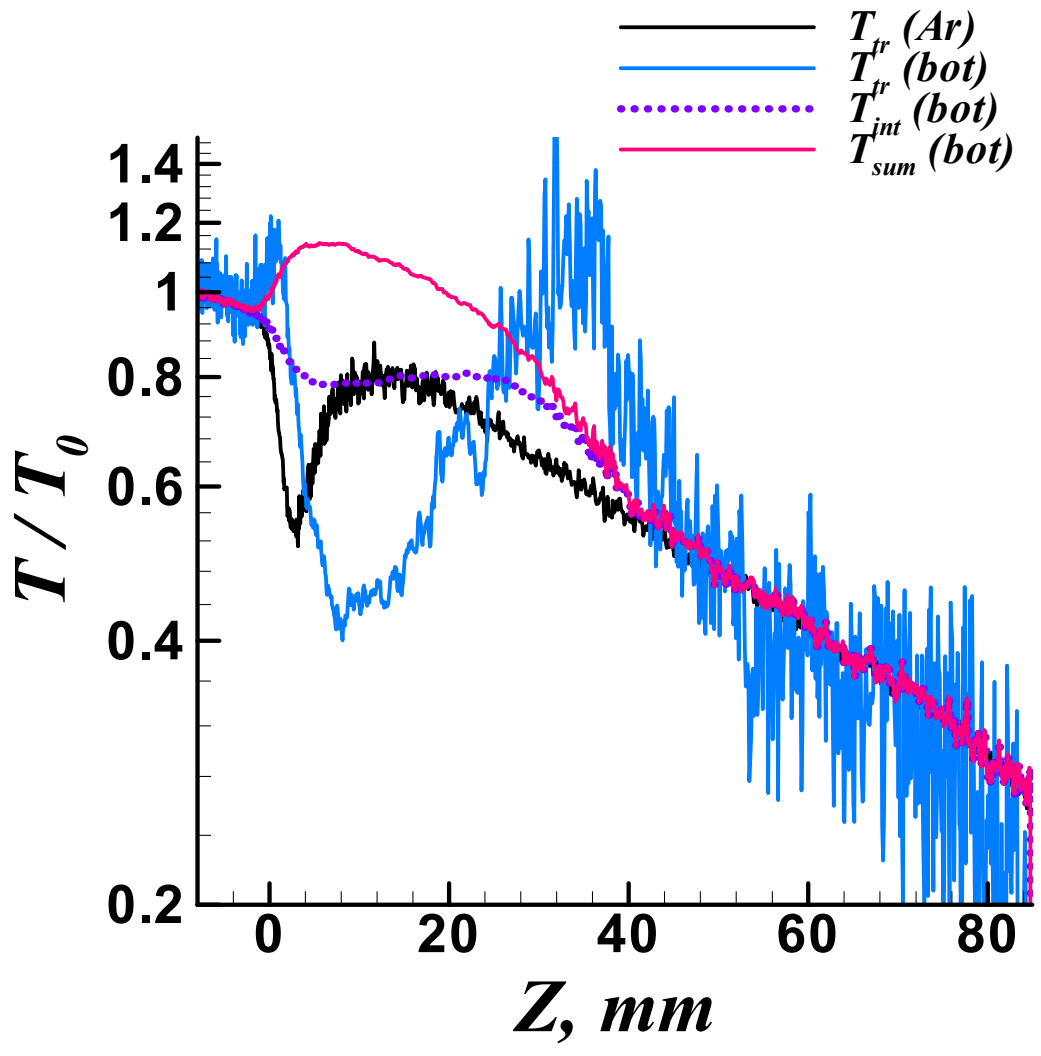

**Рис. 12-4**. Распределение температур аргона и нанокластеров (64 атома) вдоль оси струи

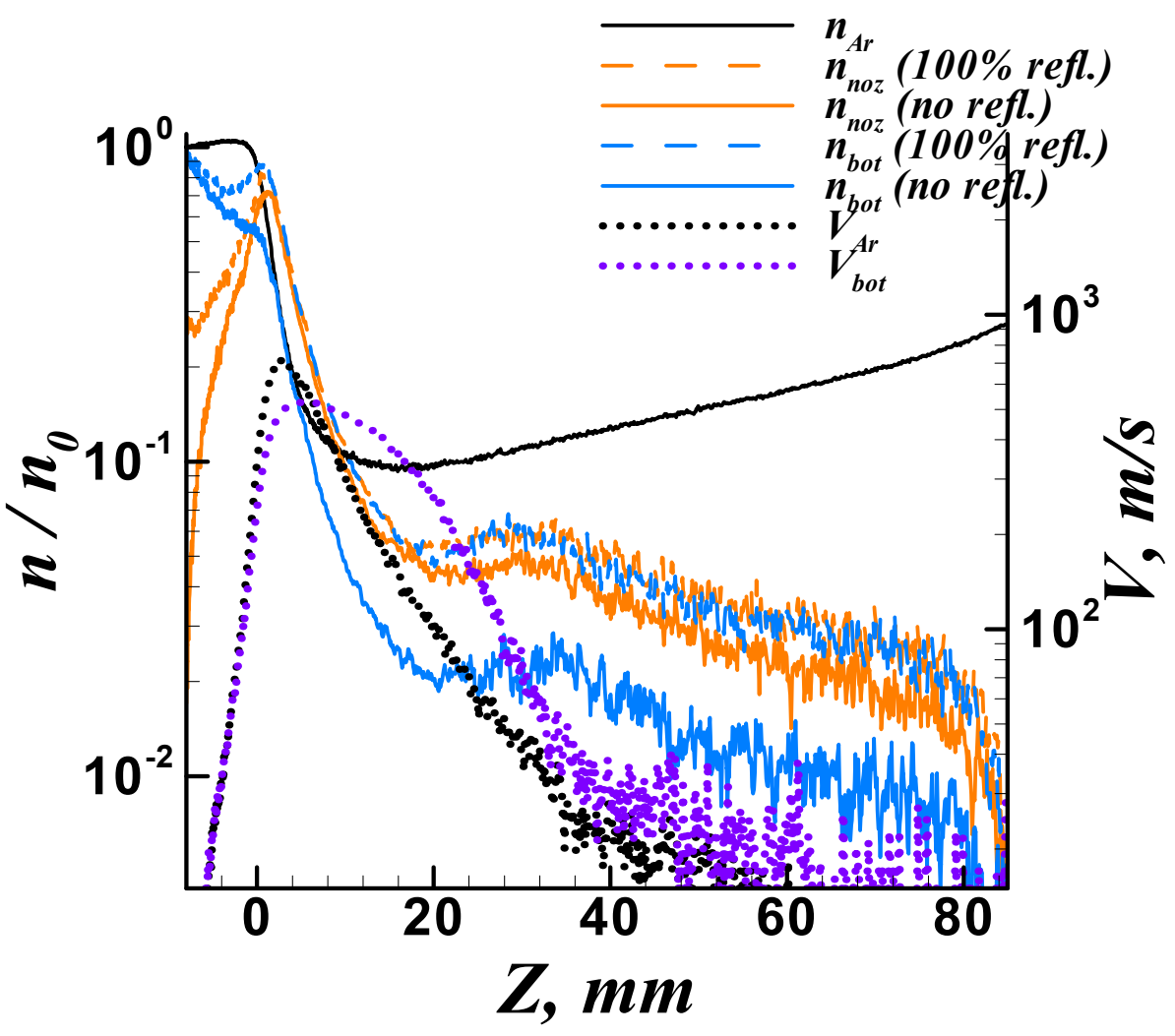

**Рис. 12-5**. Распределение плотности и скорости аргона и нанокластеров (16 атома) вдоль оси струи.

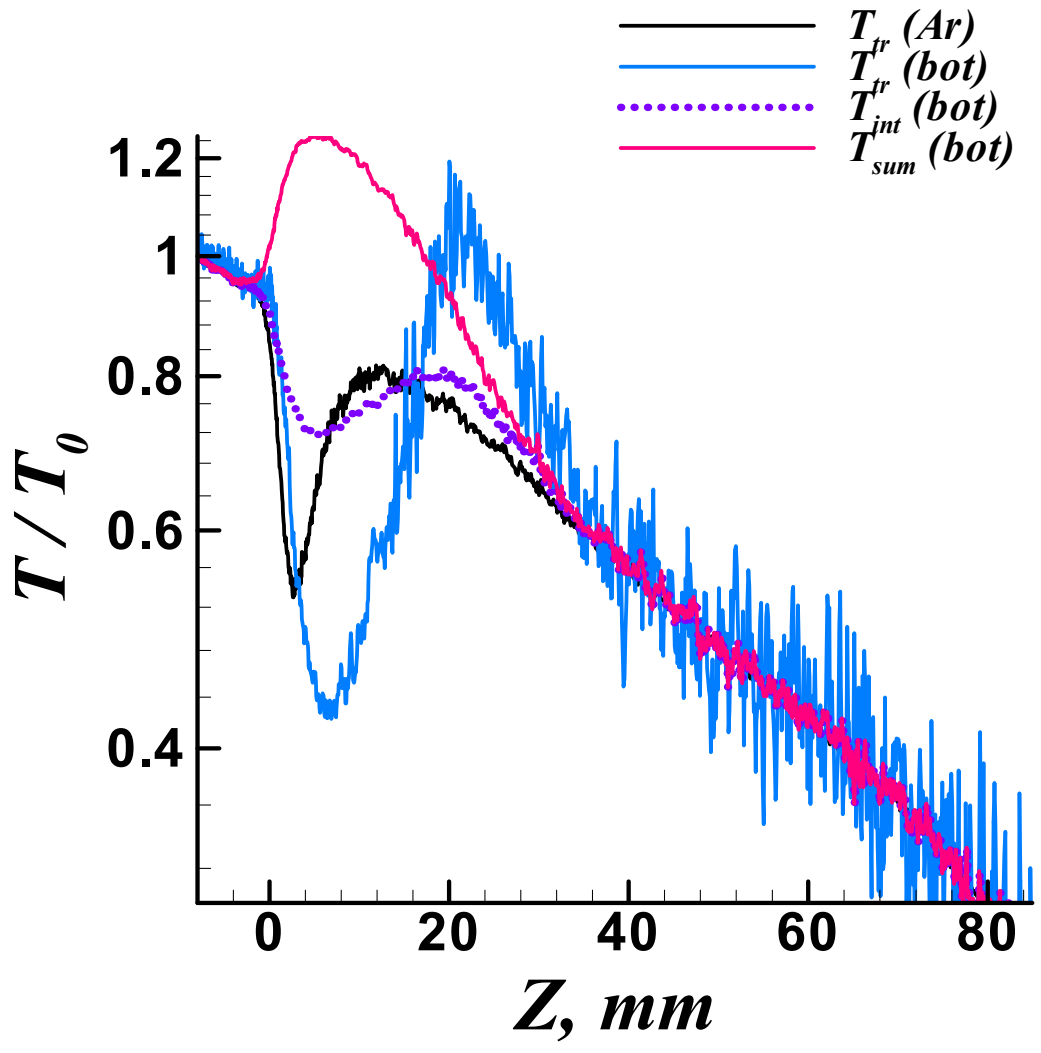

**Рис. 12-6**. Распределение температур аргона и нанокластеров (16 атома) вдоль оси струи

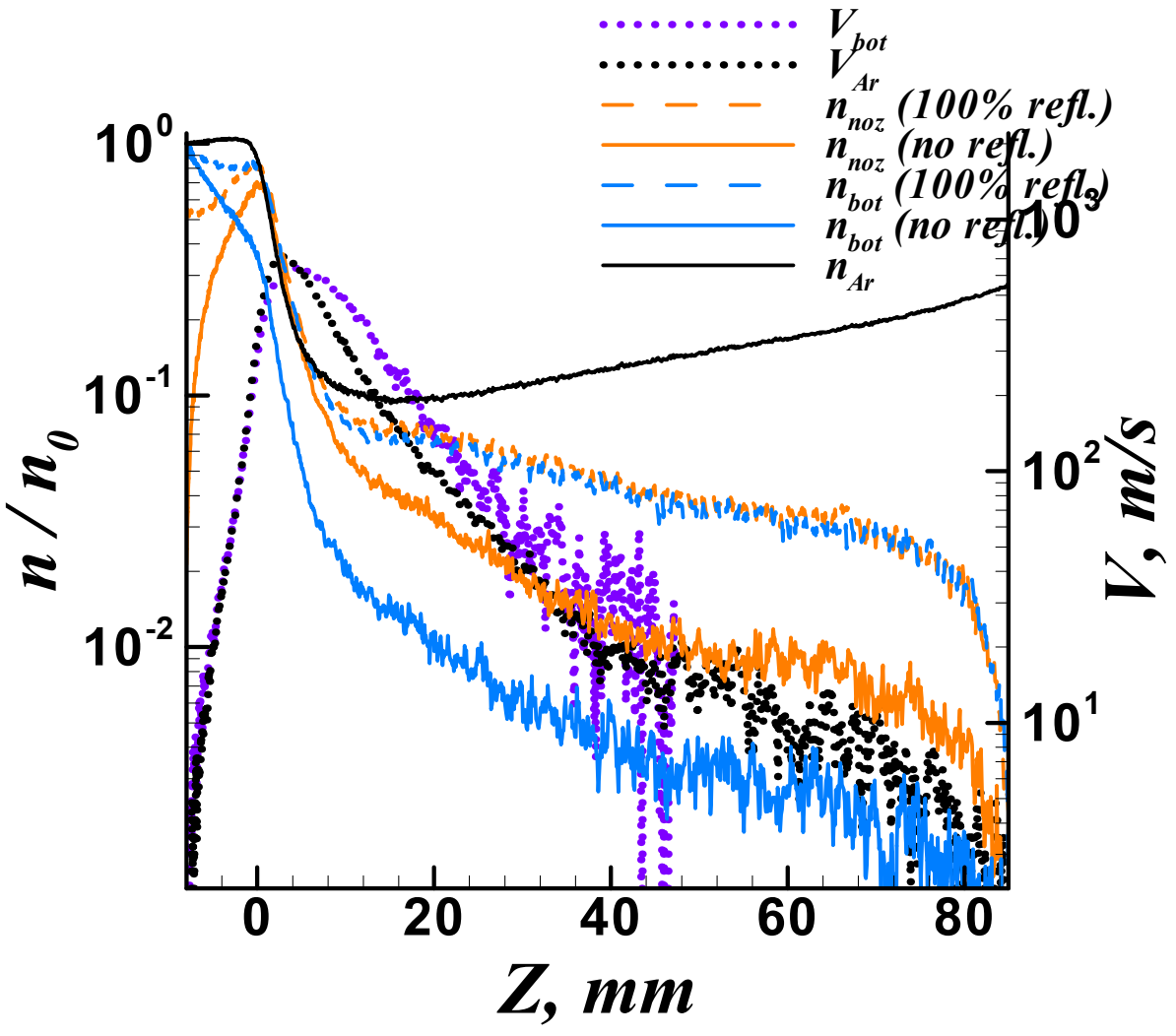

**Рис. 12-7**. Распределение плотности и скорости аргона и нанокластеров (4 атома) вдоль оси струи.

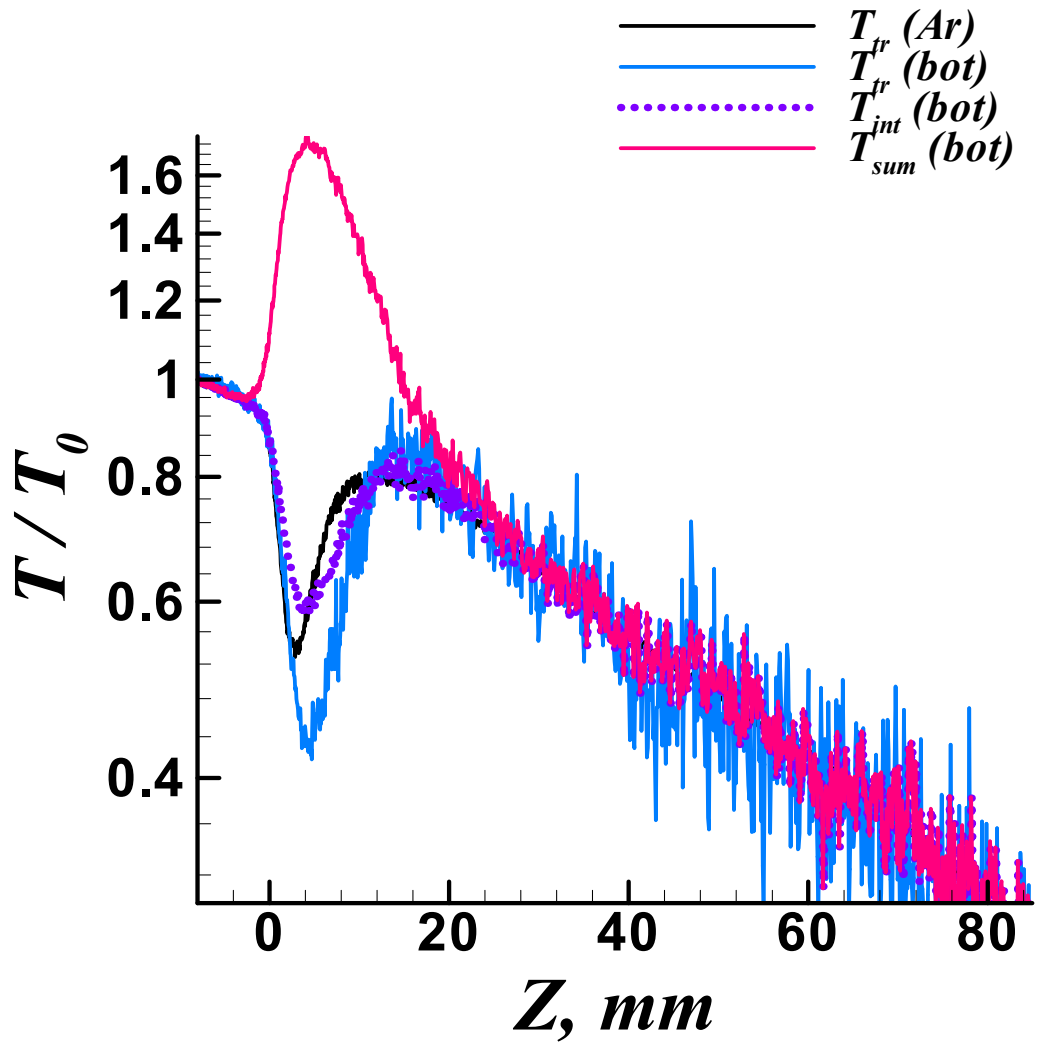

**Рис. 12-8**. Распределение температур аргона и нанокластеров (4 атома) вдоль оси струи

К сожалению, статистическое качество этих результатов не такое хорошее, как хотелось бы, так как получены они были в то время, когда как радиальная схема на основе временных множителей, так и подход для подавления флуктуаций числа частиц ещё не были доступны.

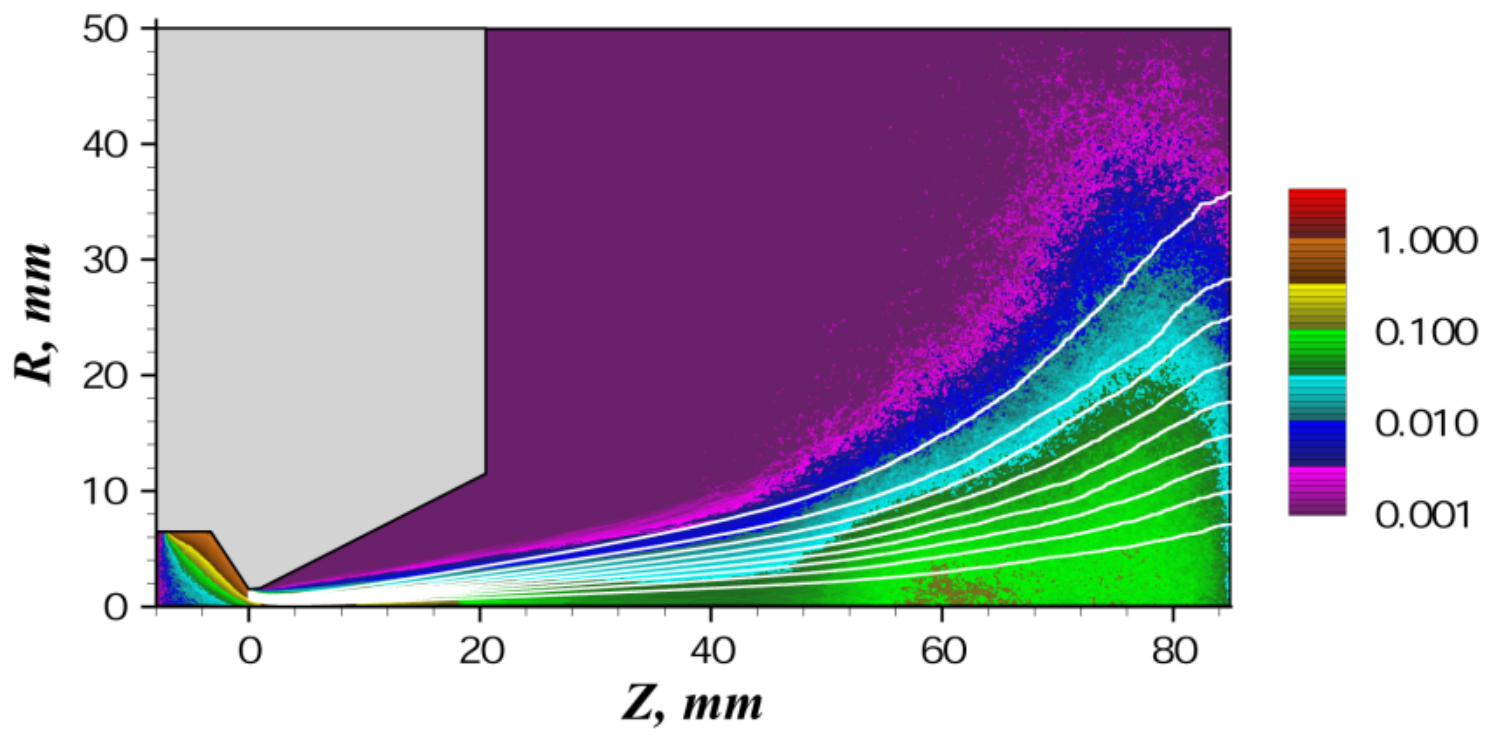

**Рис. 12-9.** Поле плотности и линии тока нанокластеров массой 256 атомов.

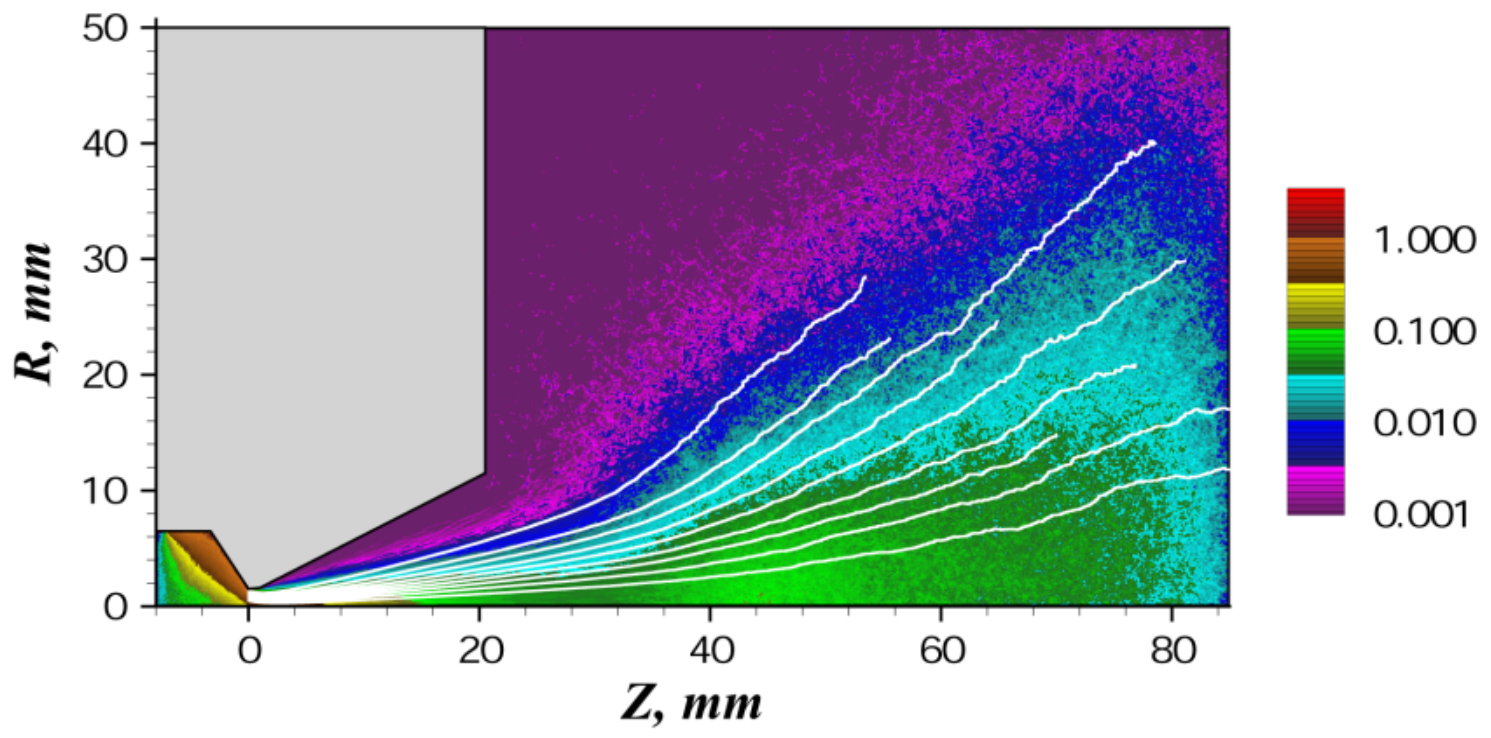

**Рис. 12-10.** Поле плотности и линии тока нанокластеров массой 64 атома.

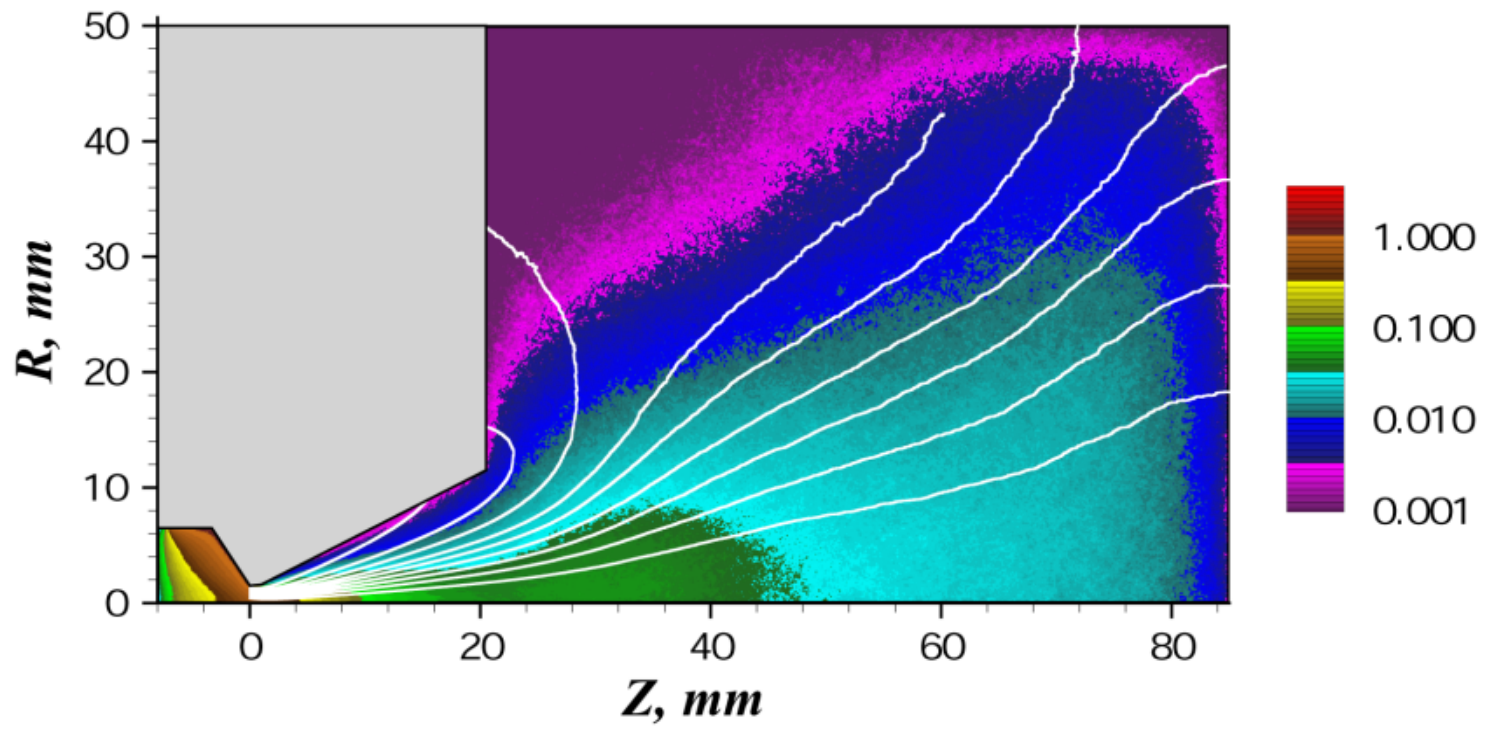

**Рис. 12-11.** Поле плотности и линии тока нанокластеров массой 16 атомов.

На **Рис. 12-9** – **Рис. 12-11** представлены линии тока на фоне поля плотности для нанокластеров 256, 64, 16 атомов, образованных на поверхности конфузора, при полном поглощении на стенках. Линии тока выбраны таким образом, чтобы в критическом сечении расход между ними был близок к постоянному. Обобщая эти данные, можно подтвердить рассуждения о двух зонах с различными режимами течения нанокластеров – направленным и диффузионным. При переходе к диффузионному режиму, в области остановки нанокластеров, направленная струя преобразуется в рассеивающееся «облако» заторможенных кластеров. Характерный размер «облака» значительно больше характерного размера ядра струи, при этом плотность в нём выше.

Для кластеров массой 16 атомов, переход к диффузионному режиму происходит вблизи среза сопла. По линиям тока видно, что значительная часть кластеров (более 30 %) поглощается на тепловом экране и на поверхности диффузора сопла.

# 13. Движение нанокластеров при наличии подложки-мишени

Для того, чтобы проанализировать результаты эксперимента по нанесению покрытия на полоску нержавеющей стали (Рис. 11-1), была проведена серия расчетов движения нанокластеров массами 4, 16, 64, 256 и 1024 в потоке аргона при условиях эксперимента по осаждению на полосу нержавеющей стали. Для сохранения осесимметричности геометрии, вместо полосы был взят диск диаметром 23 мм, чтобы попытаться сохранить число Рейнольдса. Особое внимание уделялось определению доли нанокластеров, достигших лицевой и тыльной поверхностей мишени. Все достигшие мишени нанокластеры поглощались. Рассматривались как случай образования нанокластеров внутри реактора, так и случай образования на поверхности диффузора.

Для нанокластеров, образованных на конфузоре, поля плотностей и линии тока для режимов 64, 256, 1024 приведены на **Рис. 13-1** – **Рис. 13-3**. В целом, наблюдается следующая тенденция. Тяжёлые кластеры (1024 атома) не успевают тормозиться до тепловой скорости, поэтому достигают только лицевой поверхности мишени, при этом мишени достигает почти весь поток – рассеяние отсутствует, так как диффузионный разлёт не реализуется. Нанокластеры массой 256 атомов затормаживаются в непосредственной близости от мишени, и поток успевает частично рассеяться за счёт диффузии. Тем не менее, почти весь поток всё же достигает лицевой поверхности мишени, но уже не устремляется в узкое пятно, как в предыдущем случае. Некоторая небольшая доля рассеянного потока достигает тыльной стороны мишени, в то время как в предыдущем случае поток на тыльную сторону отсутствовал вовсе. Нанокластеры массой 64 атома тормозятся ещё раньше – уже на заметном удалении от мишени, поэтому поток рассеивается более активно. Теперь на лицевую сторону попадает лишь около 1/3 потока, зато увеличился поток на тыльную сторону мишени.

На **Рис. 13-4** показано радиальное распределение плотности потока нанокластеров разной массы непосредственно перед мишенью, отнесенное к плотности потока с эмитирующей поверхности. Во всех случаях отсутствует значительное изменение поведения зависимости вблизи края мишени, что говорит в пользу того, что по данным зависимостям можно судить и о плотности потока в случае реальной мишени (узкой полосы), установленной в эксперименте. Результат для нанокластеров массой 1024 атома согласуется с данными эксперимента (**Рис. 11-1**), в котором полуширина пятна составила 7-10 мм.

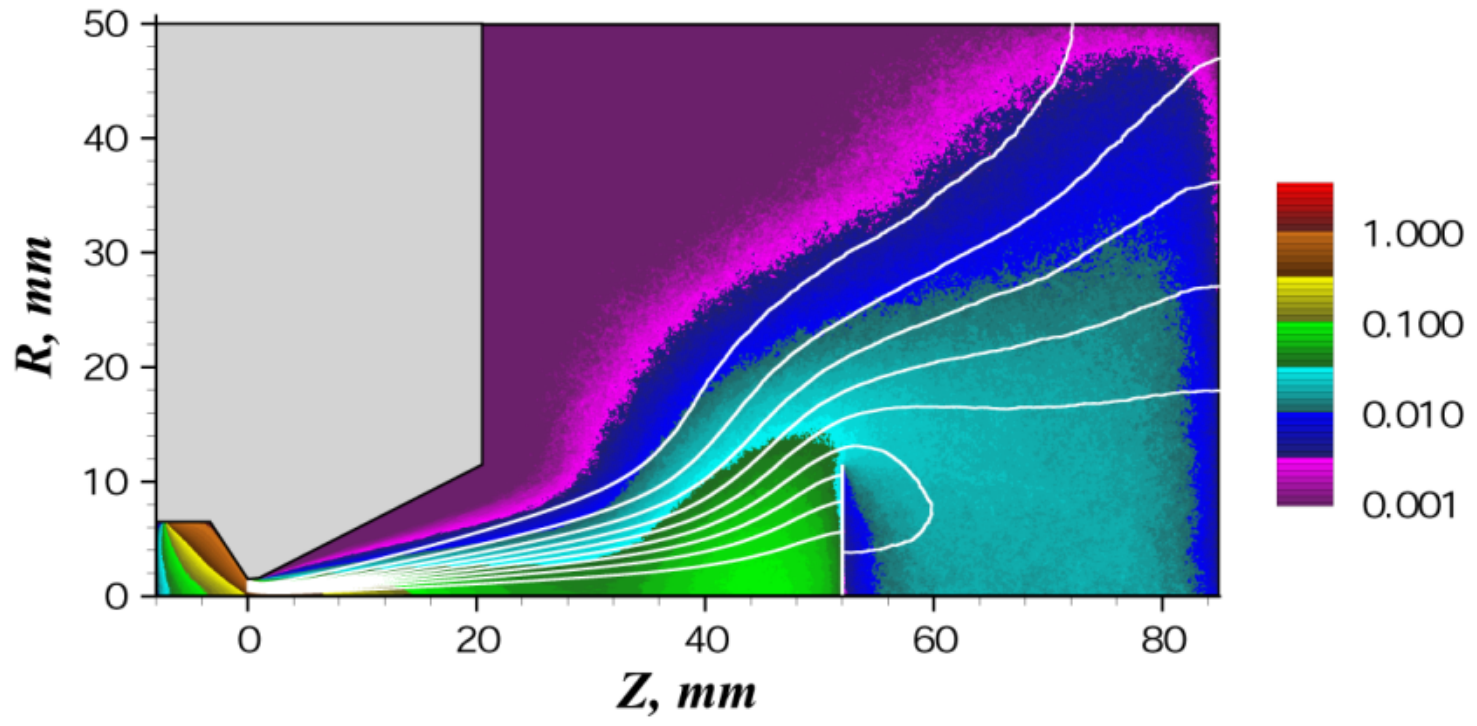

**Рис. 13-1**. Поле плотности и линии тока нанокластеров массой 64 атома, образованных внутри тигля.

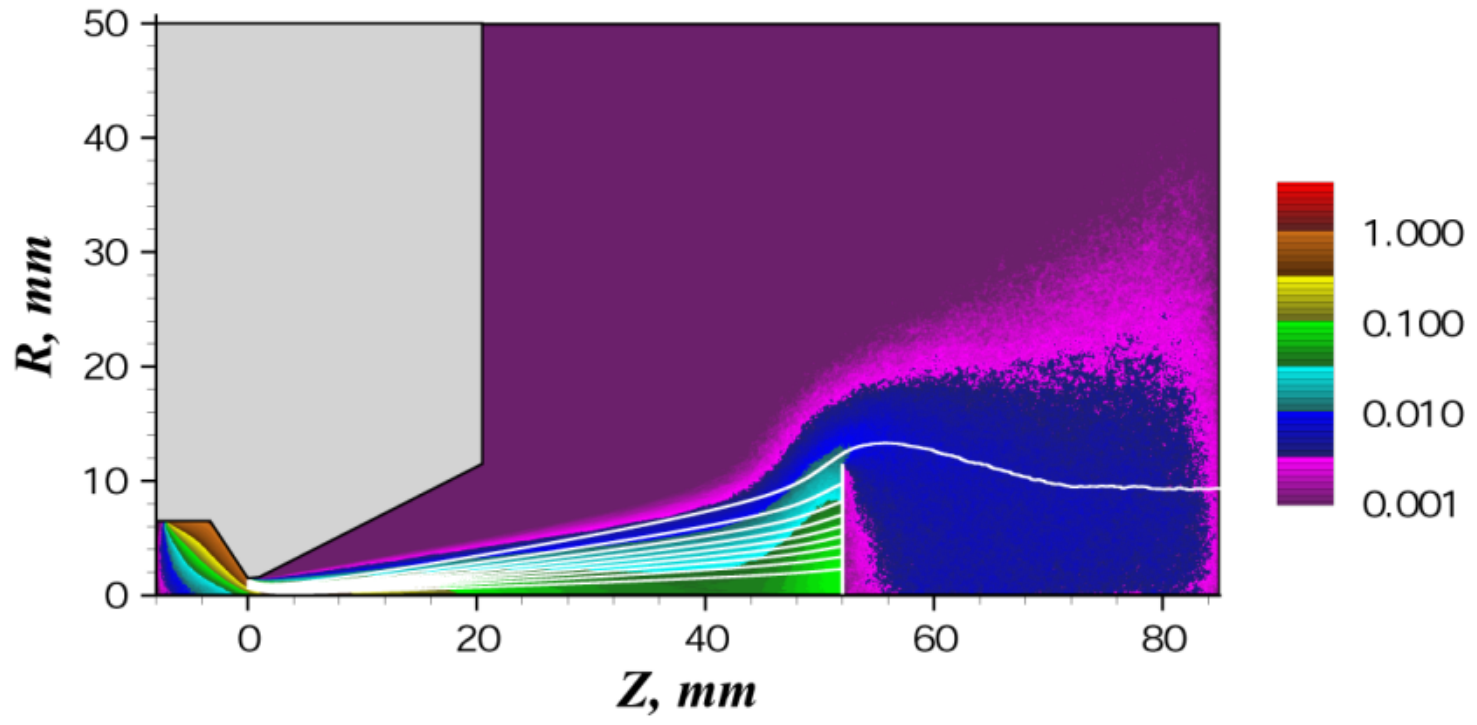

**Рис. 13-2**. Поле плотности и линии тока нанокластеров массой 256 атомов, образованных внутри тигля.

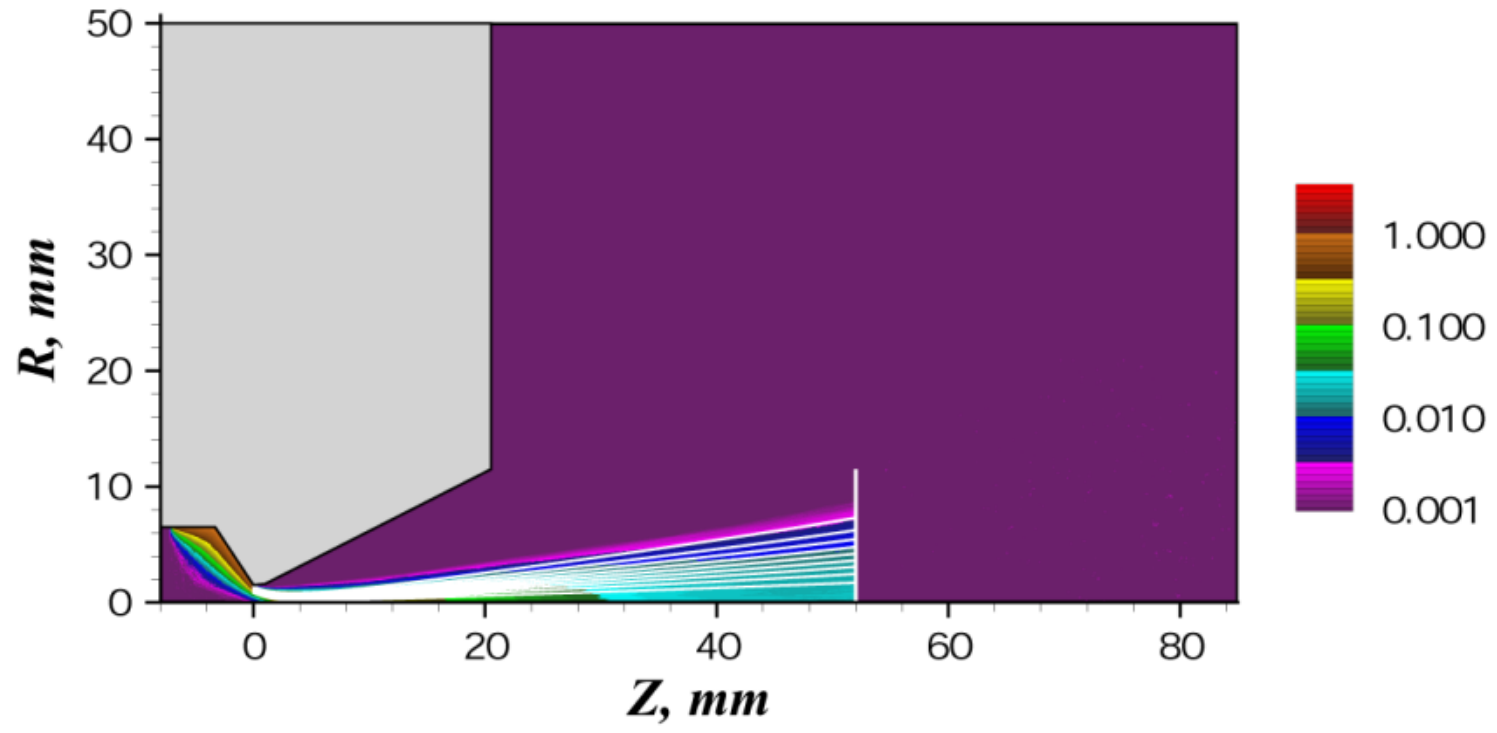

**Рис. 13-3**. Поле плотности и линии тока нанокластеров массой 1024 атома, образованных внутри тигля.

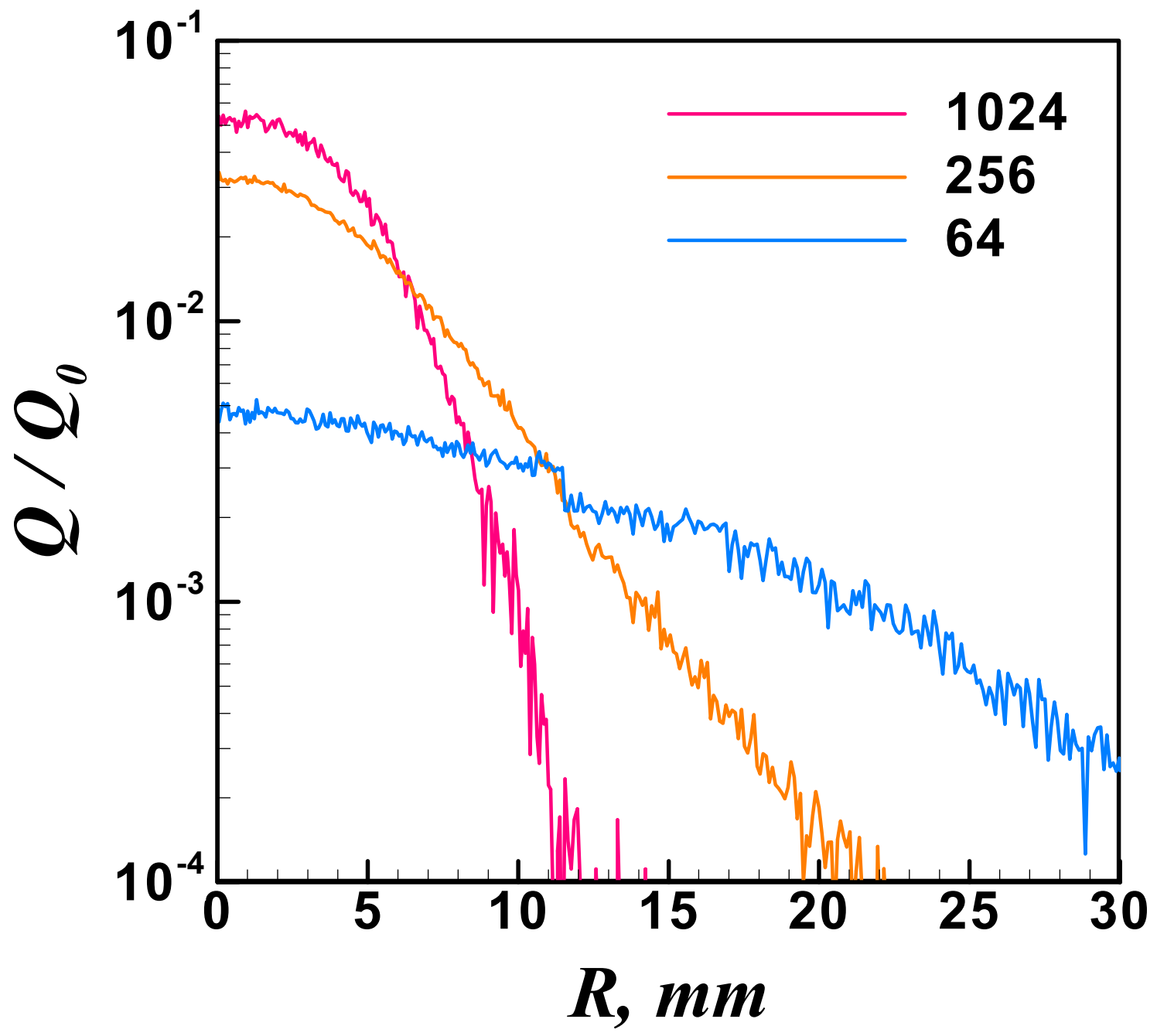

**Рис. 13-4**. Профили плотности потока нанокластеров разной массы, образованных внутри тигля, в поперечном сечении, непосредственно перед пластиной-мишенью.

Далее будет рассмотрен случай генерации нанокластеров поверхностью диффузора. Важное отличие этого случая заключено в том, что нанокластеры не имеют большой скорости направленного движения, сразу попадая в уже почти покоящийся газ пристеночного пограничного слоя. Как следствие, движение таких кластеров имеет в основном диффузионную природу.

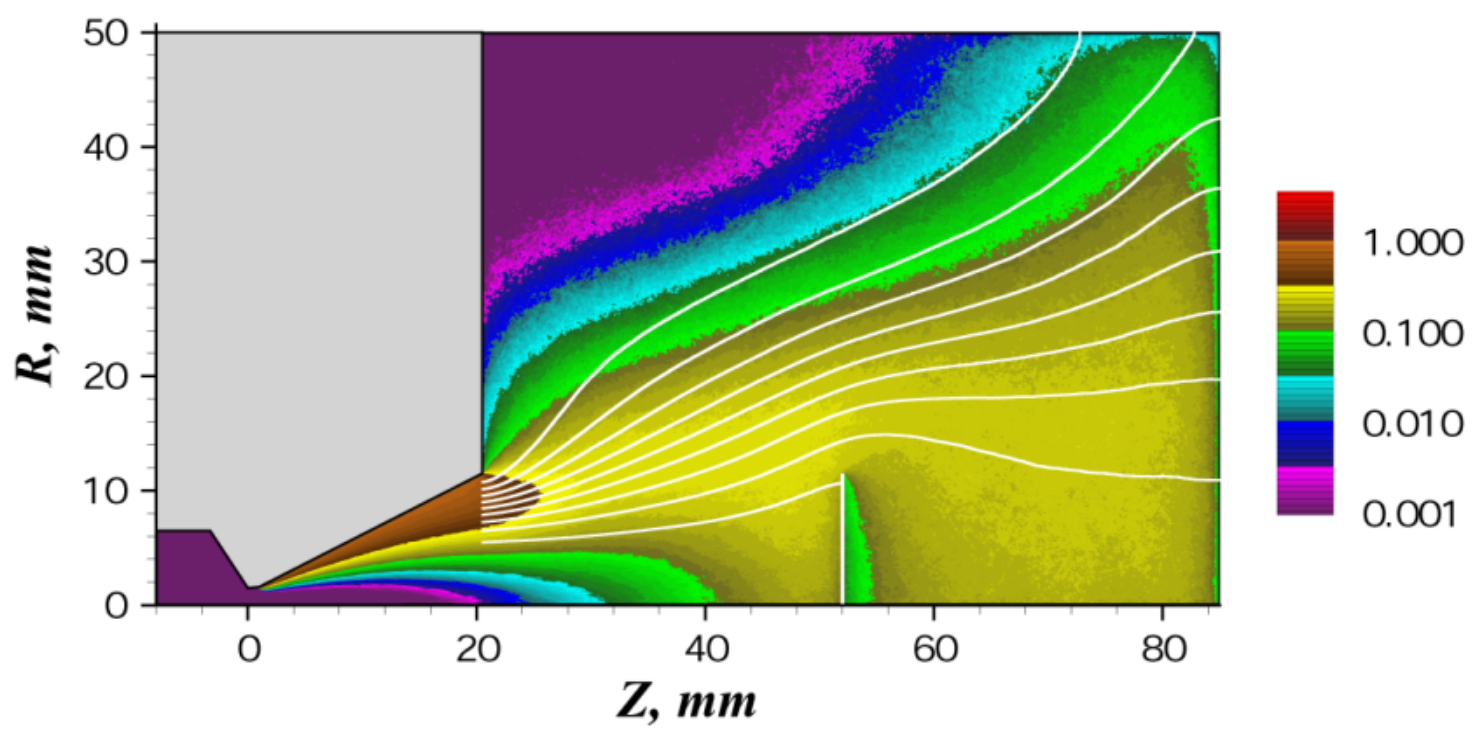

**Рис. 13-5**. Поле плотности и линии тока нанокластеров массой 256 атомов, эмитированных поверхностью диффузора.

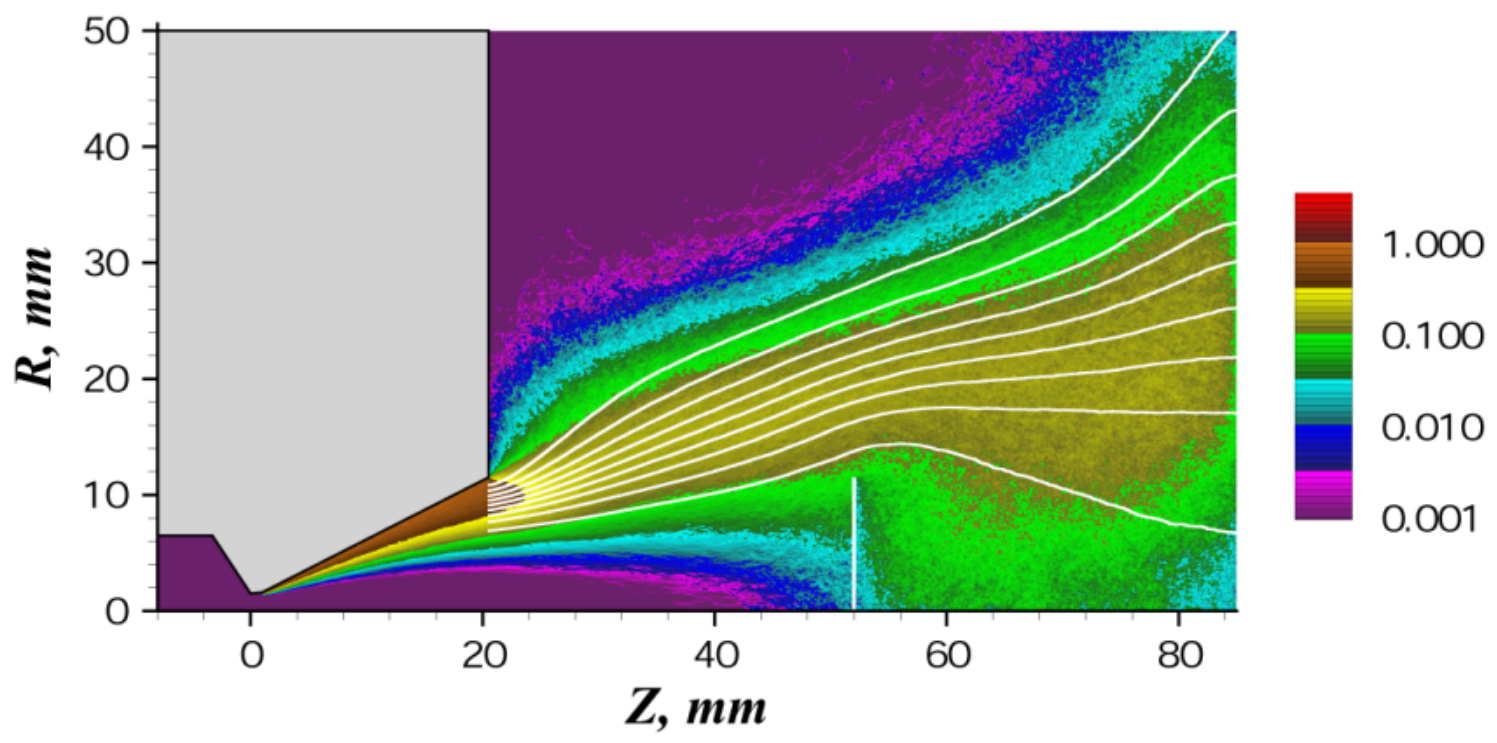

**Рис. 13-6**. Поле плотности и линии тока нанокластеров массой 1024 атома, эмитированных поверхностью диффузора.

На **Рис. 13-5** – **Рис. 13-6** представлены поля плотности и линии тока для нанокластеров массой 256 и 1024 атома. Линии тока на этот раз построены так, чтобы расход между линиями тока на срезе сопла был одинаковым. Кластеры массой 256 атомов уже довольно медленно диффундируют вглубь струи, однако вблизи мишени уже полностью заполняют ядро струи. В результате, около 10 % нанокластеров, достигших среза сопла, достигают и лицевой стороны мишени. Область за мишенью также довольно эффективно заполняется кластерами, вследствие чего можно ожидать сравнимого потока и на тыльную сторону мишени.

Кластеры массой 1024 атома ещё хуже проникают в струю, и вблизи мишени уже наблюдается значительный провал плотности вблизи оси. Однако, по-прежнему нанокластеры проникают в область за мишенью, поэтому поток на тыльную сторону мишени всё ещё сопоставим с потоком на лицевую. Этому способствует и перемешивающее воздействие вихревого движения за пластиной. Экстраполируя наблюдаемую зависимость, можно заключить, что достаточно тяжелые наночастицы (10 нм и более) практически перестанут проникать в струю и будут двигаться узким слоем вдоль продолжения конуса диффузора. Следовательно, попадание таких наночастиц в центр мишени исключено.

Полученные результаты для всех масс сведены на **Рис. 13-7**. Поток нанокластеров с поверхности конфузора на лицевую поверхность мишени монотонно растёт, выходя на плато. Поток на тыльную сторону мишени сперва растёт, достигает максимума для нанокластеров массой 64 атома, затем снижается вплоть до нуля. Отношение потоков (расстояние двумя между кривыми) всё время монотонно растёт, с ускоряющимся темпом. Для нанокластеров, образованных на поверхности диффузора, отношение потоков на лицевую и тыльную сторону мишени меняется медленно и незначительно. При этом, потоки сначала немного растут (за счёт того, что всё меньше нанокластеров осаждается на тепловом экране), затем начинают снижаться, по мере того как нанокластеры всё хуже проникают в ядро струи.

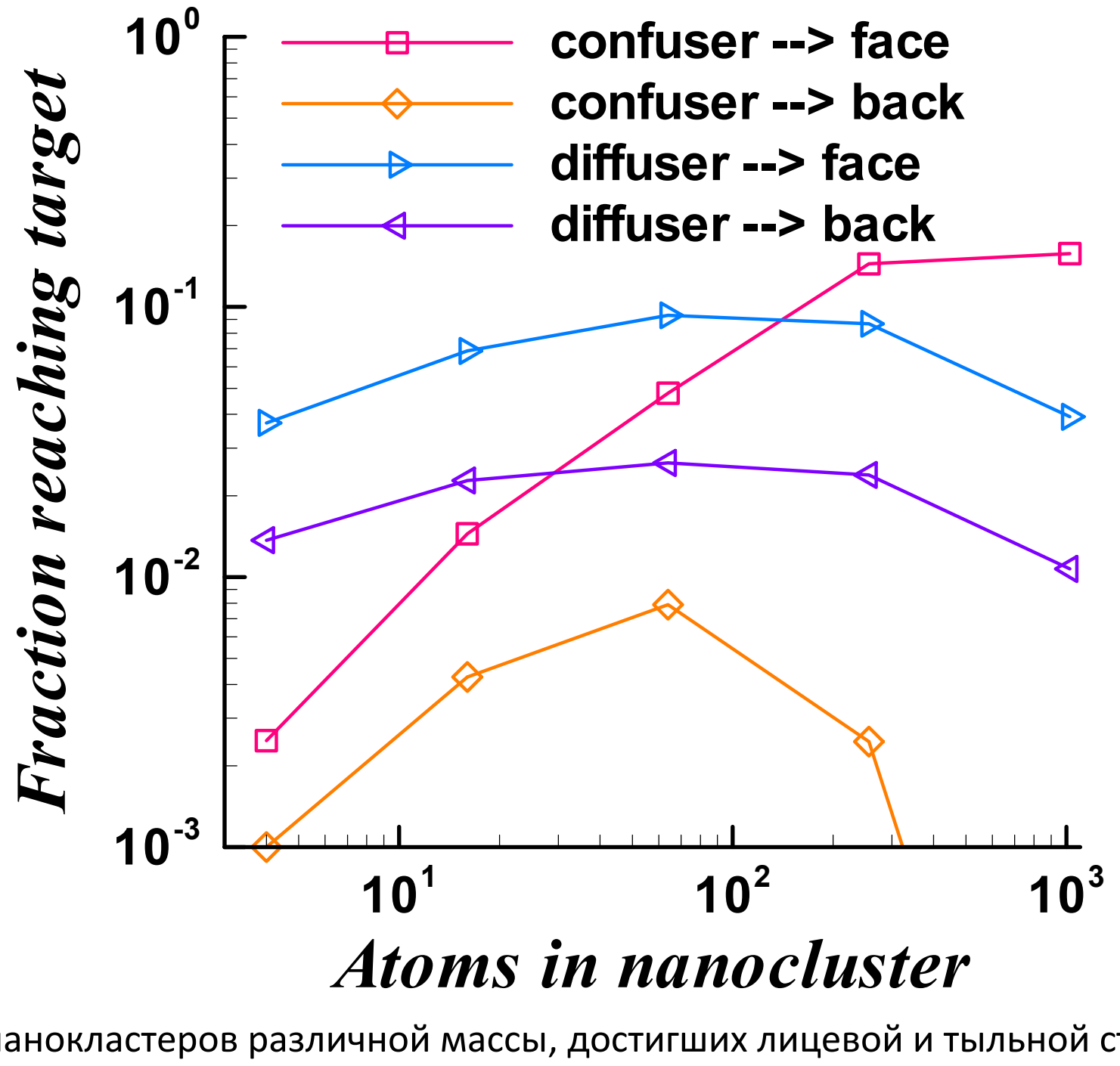

**Рис. 13-7**. Доли нанокластеров различной массы, достигших лицевой и тыльной сторон мишени, образованных как внутри тигля и на диффузоре сопла.

# 14. Выводы

Проведенный анализ газодинамики на основе численного моделирования позволяет сделать следующие выводы:

- Течения, образованные и звуковым, и сверхзвуковым соплами, формируют практически одинаковые струи ускоренных нанокластеров, если последние образованы в камере источника и имеют массу от 64 атомов.
- Диффузор сверхзвукового сопла, однако, может выступать поглотителем паров серебра и существенно влиять на концентрацию паров в камере расширения.
- Потоки серебряного пара на лицевую и тыльную стороны мишени сопоставимы.
- Нанокластеры массой вплоть до 1024 атомов, образованные на поверхности диффузора, также дают сопоставимые потоки на лицевую и тыльную стороны мишени.
- Нанокластеры массой более 1024 атомов, образованные на поверхности диффузора, с ростом массы, всё меньше рассеиваются, всё хуже проникают в ядро струи и, как следствие, должны осаждаться на лицевой поверхности мишени кольцом.
- Для нанокластеров, образованных внутри камеры источника, отношение их потоков на лицевую и тыльную стороны мишени быстро нарастает с ростом их массы. Нанокластеры массой от 1024 не попадают на тыльную сторону мишени.
- Полученная численно плотность потока нанокластеров серебра массы 1024 атома на подложку хорошо согласуется с экспериментальными данными по осаждению на узкую пластину из нержавеющей стали.
- Конкретное место образования нанокластеров внутри камеры источника слабо влияет на параметры сформированной струи ускоренных нанокластеров.
- Основная масса серебряного покрытия происходит из нанокластеров массой более 256 атомов, образованных внутри камеры источника.
- Нанокластеры массой более 256 атомов, образованные внутри камеры источника и ускоренные до высокой скорости, могут иметь сравнительно высокую энергию при подлёте к подложке.
- Поток паров серебра на мишень не вносит значительного вклада в формирование центров конденсации на мишени, но, возможно, может способствовать к дальнейшему росту осаждённых нанокластеров на её поверхности.
- Предсказана (и, впоследствии, подтверждена экспериментально) сильная неизотермичность внутренних поверхностей источника серебра, в частности, приводящая к конденсации серебра на конфузоре сопла. Это свидетельствует о том, что нанокластеры образуются гетерогенно на поверхностях камеры источника.
- Переходные режимы источника, при которых аргон не подаётся, характеризуются ещё более высокими пересыщениями пара на внутренних недогретых поверхностях источника. После подачи аргона, пересыщение заметно падает. Возможно, это оказывает влияние на процессы образования нанокластеров на этих поверхностях.
- Численное исследование усовершенствованным методом ПСМ позволило сделать дополнительные выводы о происходящих процессах, недоступные при непосредственном анализе экспериментальных данных.

В качестве практических рекомендаций, полезных для дальнейших экспериментальных исследований, можно заключить следующее:

- Зародыши осаждаемых кластеров формируются гетерогенным образом на внутренних недогретых поверхностях камеры источника. Как следствие, оптимизация источника для более эффективной генерации зародышей осаждаемых частиц неизбежно связана с изучением и оптимизацией, прежде всего, именно гетерогенных процессов внутри источника.

- Варьирование давления несущего газа в источнике и фонового газа в камере расширения позволяет управлять энергией и составом долетающих до мишени нанокластеров.

- Энергия нанокластеров, подлетающих к подложке, может быть достаточно велика для химического взаимодействия с полимером при осаждении металлополимерных нанокомпозитов. Что, в частности, может оказывать влияние на бактерицидные свойства полученного покрытия серебро+политетрафторэтилен.

- Пары серебра не вносят значимого вклада в формирование центров конденсации, но, возможно, способствуют росту размеров осаждённых на поверхности частиц серебра. Это предположение имеет смысл изучить экспериментально. Диффузор сопла может быть задействован как регулятор концентрации паров серебра в камере расширения.

# 15. О гетерогенных процессах образования кластеров

Проведенное численное исследование позволило сделать однозначный вывод: нанокластеры, являющиеся зародышами конденсации на мишени, образуются гетерогенным образом на внутренних поверхностях камеры источника. Это возможно в силу того, что источник паров серебра неизотермичен, и его недогретые поверхности находятся в атмосфере пересыщенного пара.

Подробный анализ гетерогенных процессов зародышеобразования лежит за пределами охвата настоящей работы, посвященной газодинамическим процессам. Хорошим источником информации по этой теме является монография (3).

Некоторые возможные механизмы образования зародышей конденсации:

- Флуктуационный механизм поверхностного зародышеобразования из адсорбированных паров с последующей десорбцией надкритических зародышей.
- Гомогенное зародышеобразование в газовой фазе.
- Эмиссия поверхностью твёрдых частиц (например, окислов металлов), которые могут служить впоследствии центрами конденсации.
- Образование на поверхности и последующее реиспарение пористого конденсата. По мере ухудшения проникновения пара вглубь конденсата, за счёт реиспарения, частицы конденсата могут терять связь с поверхностью.

Гетерогенное образование зародышей может быть на десяток порядков эффективнее гомогенного зародышеобразования, что также показано в (3). На основании этого делается вывод, что именно гетерогенные процессы перспективны для технологических приложений.

Сильное влияние на процесс конденсации могут оказывать неконденсирующиеся примеси. Коагуляция частиц также может влиять на структуру конденсата.

Бимодальные функции распределения по размерам частиц конденсата, полученного на поверхности-мишени, характерны для процессов осаждения при больших пересыщениях и наличии дополнительных возмущающих факторов.

Данный раздел не претендует на конструктивный анализ поверхностных эффектов в источнике паров серебра. Он лишь ещё раз обращает внимание на то, что оптимизация газодинамических процессов в источнике далеко не исчерпывает собой все возможности повышения производительности источника. Изучение поверхностных эффектов внутри источника представляет собой ещё больший интерес и является довольно широким полем для дальнейших исследований.

## 16. Благодарности

Автор выражает благодарность Институту Теплофизики СО РАН за поддержку при выполнении настоящей работы. Особая благодарность Алексею И. Сафонову за сотрудничество и предоставление экспериментальных результатов.

## Список литературы